\documentclass[aps,prd,twocolumn,superscriptaddress,preprintnumbers,floatfix,nofootinbib,notitlepage,showkeys,showpacs]{revtex4-1}

\usepackage[utf8]{inputenc}

\usepackage{graphicx}
\usepackage{hyperref}
\usepackage{latexsym}
\usepackage{amsmath}
\usepackage{amssymb}
\usepackage{bbm}
\usepackage{ulem}
\usepackage{pdfsync}
\usepackage{epsfig}
\usepackage{epstopdf}
\usepackage{subfigure}
\usepackage{xcolor}
\usepackage{comment}
\usepackage{slashed}
\usepackage{multirow}
\usepackage{xfrac}
\usepackage{enumerate}

\newcommand{\avg}[1]{\langle #1 \rangle}

\begin{document}

\title{Uncertainties in the supermassive black hole abundance \\ and implications for the GW background}

\author{Gabriela Sato-Polito}
\email{gsatopolito@ias.edu}
\affiliation{School of Natural Sciences, Institute for Advanced Study, Princeton, NJ 08540, United States}

\author{Matias Zaldarriaga}
\affiliation{School of Natural Sciences, Institute for Advanced Study, Princeton, NJ 08540, United States}

\begin{abstract}
The present-day mass function of supermassive black holes is the most important observable quantity for the prediction and theoretical interpretation of the gravitational wave background (GWB) measured by pulsar timing arrays (PTAs). Due to the limited sample size of galaxies with dynamically inferred SMBH masses, more readily measurable galaxy properties $X$ that correlate with the black hole mass are used as labels (via scaling relations $M_{\bullet}-X$), which can then be counted in a larger galaxy catalog to produce a measurement of the mass function. Estimating the amplitude of the GWB from the local mass function is therefore simpler than general measurements of scaling relations and galaxy mass/luminosity functions for two reasons: the contribution to the characteristic strain is dominated by a narrow range of masses, and the mass proxy $X$ is always marginalized over. While consistent errors in $X$ in both catalogs are irrelevant, relatively small biases between them can produce significant shifts in the predicted SMBH abundance. In this work, we explore measurements of the SMBH mass function using different mass proxies through a set of catalogs with a number of redundant measurements between them. This enables us to investigate internal inconsistencies that lead to discrepancies in the final black hole abundance, while minimizing observational systematic biases induced by combining disparate sets of measurements. We focus on 3 proxies: the velocity dispersion $\sigma$, K-band luminosity $L$, and a combination of $L$ and radius $R$ defined by the fundamental plane. We show that all three can be reconciled to some degree, but highlight the remaining dependence on poorly-quantified systematic corrections between the scaling relation catalogs and the mass function catalogs, as well as the potential impact of selection effects.

\end{abstract}

\maketitle

\section{Introduction}\label{sec:intro}
All major pulsar timing array (PTA) collaborations --- the European PTA and Indian PTA (EPTA+InPTA), North American Nanohertz Observatory for Gravitational waves (NANOGrav), Parkes PTA (PPTA), and the Chinese PTA~\cite{EPTA, NANOGrav_stoc, PPTA, CPTA}--- recently announced first evidence for gravitational waves (GWs) with varying degrees of significance. Further studies of these early results have demonstrated consistency between them, with $<1\sigma$ discrepancies despite observational and modeling differences~\cite{IPTA}. More recently, the MeerKAT PTA \cite{MPTA} also reported a common red noise signal across pulsars. Determining the origin of the $\sim$nHz GW signal is therefore a crucial question that will remain pressing as PTA sensitivities continue to improve.

The most likely candidates are inspiralling supermassive black holes (SMBHs; see, e.g., Ref.~\cite{Burke-Spolaor:2018bvk}). All galaxies appear to host SMBHs at their centers~\cite{1995ARA&A..33..581K, 1998Natur.395A..14R} and are expected to merge following the merger of the host galaxies~\cite{1980Natur.287..307B}. Observations in our local universe yield the most precise and direct measurements of SMBHs. Their masses can be directly inferred from stellar or gas kinematics (see, e.g., Refs.~\cite{kormendy_ho, mcconnell_ma}), along with various properties of the host galaxies which are found to correlate with black hole mass. These correlations suggest a shared evolution between SMBHs and host galaxies and offer valuable insight into the evolution of galaxies over their lifetimes. Furthermore, combining such scaling relations, which are typically measured in nearby ($\lesssim 100$~Mpc) samples of $\sim 100$ galaxies (BH catalog), with a larger galaxy catalog from which mass or luminosity functions can be inferred (MF catalog) allows one to estimate the local abundance of SMBHs~\cite{2004MNRAS.351..169M, vika2009, shankar2009, shankar2013}.

The amplitude of the GWB produced by merging black holes can be directly related to the mass function of the remnant population today~\cite{Phinney:2001di}. This simple relation provides a robust estimate for the signal expected in PTA measurements that is fairly insensitive to the details of the merger history, as well as a strict upper limit for the amplitude (explored in detail in Refs.~\cite{Sato-Polito:2023gym, Sato-Polito:2024lew, Sato-Polito:2025ivz}). Curiously, we showed in Ref.~\cite{Sato-Polito:2023gym} that most estimates of the SMBH mass function lead to an underprediction of the GW background compared to PTA measurements. The later results of Ref.~\cite{Sato-Polito:2024lew} suggest that boosts to the mass function at the massive end ($\log_{10} M_{\bullet}/M_{\odot} \gtrsim 10$) would lead to evidence for Poisson fluctuations in the PTA spectrum that are currently disfavored in the NANOGrav 15yr free spectrum~\cite{NANOGrav_stoc, Lamb_free_spec}, thus suggesting a resolution at lower masses.

There is a long history of debate regarding both the scaling relation $M_{\bullet}-X$, which relates the black hole mass $M_{\bullet}$ to a galaxy property $X$, measurements of the abundance $\frac{dn}{dX}$ of galaxies per mass proxy $X$, and the ultimate impact on the inferred SMBH mass function. For example, Refs.~\cite{2007ApJ...662..808L, tundo2007, 2007ApJ...660..267B, 2016MNRAS.460.3119S} discuss discrepancies between black hole masses inferred via the velocity dispersion $\sigma$ of the host galaxy and via luminosity $L$ or stellar mass $M_{*}$ (see Appendix~A of Ref.~\cite{Shankar:2025unm} for a comprehensive summary of this discussion), while a number of studies have found differences in luminosity and stellar mass functions~\cite{Bernardi_LMfunc_2013, DSouza_2015, Leja_2020, Liepold_2024} (although there appears to be fewer disagreements regarding the velocity dispersion function, e.g. \cite{bernardi_MF}). 

This work aims to clarify the most significant sources of uncertainty to the black holes mass function estimate, particularly in the regime relevant to predictions of the GW background amplitude. There are two features that simplify the task of estimating the strain amplitude, relative to general studies aiming to measure scaling relations and mass functions independently: 
\begin{enumerate}
    \item that it is dominated by a fairly narrow range of (high) SMBH masses,
    \item and, since the mass function is inferred via
    \begin{equation}
    \frac{dn}{d\log M_{\bullet}} = \int dX \ p(\log M_{\bullet}|X) \frac{dn}{dX},
    \label{eq:BHMF}
\end{equation}
where $p(\log M_{\bullet}|X)$ is the scaling relation, this means that the mass proxy $X$ is always marginalized over.
\end{enumerate}
Point (1) restricts the parameter range of galaxies that are of consequence and (2) clarifies possible sources of errors. The mass proxy $X$ can be thought of as a label applied to galaxies in the BH sample, which is then used to count in the MF sample. Consistent shifts/errors across both catalogs do not bias the inference of the BH mass function and, in principle, any scaling relation should result in the same mass function. The fact that this is not satisfied in practice can mean:
\begin{enumerate}[(i)]
    \item the presence of systematic errors that lead to relative shifts in the property $X$ that would be assigned to the same galaxy in the BH and MF samples,
    \item that the BH catalog is biased relative to the MF catalog,
    \item or mismodeling of the scaling relation.
\end{enumerate}
Point (i) may be due to systematics in the photometry, which include sky subtraction, modeling of the surface brightness profile and extrapolation to the total luminosity, as well as the conversion from light to mass in the stellar mass case \cite{Bernardi_LMfunc_2013}, and may also include inconsistent conventions between the two catalogs. Point (ii) has been the subject of much debate and may be driven by selection based on the ``sphere of influence'' $r_i=GM_{\bullet}/\sigma^2$ \cite{2010ApJ...711L.108B, 2016MNRAS.460.3119S, 2023MNRAS.518.1352S, Shankar:2025unm}, although the central surface brightness also poses limitations \cite{MASSIVE_2014}. Both (i) and (ii) manifest as inconsistencies in the relation between galaxy properties across catalogs. Finally, (iii) refers to the model for $p(\log M_{\bullet}|X)$ being insufficient to capture the observed galaxy distribution, such as fitting a single power law and scatter to a distribution with different populations, or the need for a quadratic term. In general, if (iii) is a significant factor, it should be fairly self-evident.

The goal of this work is to study some of these possibilities using the 2MASS~\cite{2006AJ....131.1163S} and 6dFGS~\cite{6dFGS_DR1, 6dFGS_DR3} catalogs as testbeds. The choice of data set is motivated by the redundancy between galaxies in the BH and MF samples captured in these catalogs, in order to minimize the impact of point (i) which can be quite large. We focus on three galaxy properties: $\sigma$, total K-band luminosity $L$, and effective radius $R$. While $M_{*}$ is a commonly adopted choice of mass proxy, we opt to focus on more directly observable quantities, once again to mitigate the biases suggested by (i). However, given the relation between stellar mass and K-band luminosity, it is plausible that some of the conclusions found here may translate.

We begin with a more quantitative version of the discussion introduced here in Sec.~\ref{sec:general} and quantify the impact that offsets in the mass proxy $X$ between the two catalogs produce on the characteristic strain prediction. We then investigate the black hole mass function inferred using $\sigma$, $L$, and a virial parameter $X_{\rm vir} \equiv \log L + \alpha_{\rm vir} \log R$, as well as the internal relations between these galaxy properties in the BH and MF catalogs. If the relation between these properties were the same in both and the scaling relation is appropriately modeled, then all estimates of the mass function should also be identical. 

Using the fundamental plane sample from the 6dFGS catalog~\cite{6dFGSv_data,6dFGSv_FP}, we study the relations between $\sigma-X_{\rm vir}$ and $\sigma-L$ in the MF sample and compare it to the BH sample. We find (unsurprisingly) that they are not the same. Further investigation shows that most of the inconsistency between $\sigma-L$ in the BH and MF samples can be attributed to differences in surface brightness of the galaxies. This can be understood from the fact that $\sigma-L$ corresponds to a particular projection of the fundamental plane and does not account for the radius dependence.

We also discuss the impact of magnitude, radius, and velocity dispersion corrections that add a relative shift in the mass proxy between BH and MF samples. Putting all together, we show the resulting SMBH mass functions based on different mass proxies and with different corrections to observed galaxy properties. We find a remarkable agreement between the velocity dispersion function (VDF) measured in this work, based on the $\sigma-X_{\rm vir}$ scaling relation, and the direct measurement from SDSS ~\cite{bernardi_MF}, suggesting a robust measurement of the VDF from galaxy surveys. The only known correction to the velocity dispersion between the BH and MF samples stems from the fact that they are measured at different apertures (roughly $R_e$ for the local samples, and $R_e/8$ for the MF catalogs). Since velocity dispersion profiles for massive galaxies are consistently found to rise at the center, including such a correction strictly decreases the inferred amplitude of the BHMF. Although the magnitude and radius corrections appear to be significant and are poorly quantified, we find a reasonable agreement between these approaches.

We denote $\log_{10}$ as $\log$ throughout and adopt Planck~\cite{Planck:2018vyg} cosmological parameters for all calculations. 

\section{General considerations} \label{sec:general}
The amplitude of the SGWB can be estimated from the remnant supermassive black hole population at $z=0$ \cite{Phinney:2001di}. Following the notation and discussion in Refs.~\cite{Sato-Polito:2023gym, Sato-Polito:2024lew}, the square of the characteristic strain $h_c$ is given by
\begin{equation}
\begin{split}
    h^2_c(f) =& \frac{4 G^{5/3} f^{-4/3}}{3 \pi^{1/3} c^2} \avg{(1+z)^{-1/3}} \avg{\eta} \times \\ &\int d\log M_{\bullet}\ M_{\bullet}^{5/3} \frac{dn}{d\log M_{\bullet}},
    \label{eq:h2c}
\end{split}
\end{equation}
where $f$ is the observed GW frequency, $z$ is the redshift of the source, $M_{\bullet}$ is the total black hole mass of the binary, $\eta=q/(1+q)^2$ is the symmetric mass ratio and $q=m_1/m_2$ is the ratio of the component masses. We have defined the redshift and mass ratio averages
\begin{equation}
    \avg{(1+z)^{-1/3}} = \int dz \frac{p_z(z)}{(1+z)^{1/3}} \quad \text{and} \quad \avg{\eta}= \int dq\ p_q(q) \eta.
\end{equation}
For the redshift and mass ratio distributions, we adopt the same parametrization and fiducial values as given in Eq.~18 of Ref.~\cite{Sato-Polito:2023gym}, and refer the reader to the discussion presented there regarding the motivation for such parametrization and the impact of varying it.

In order to estimate the SGWB and, in particular, the upper limit thoroughly discussed in Ref.~\cite{Sato-Polito:2023gym}, we must therefore measure the present-day SMBH mass function. This is typically achieved by combining scaling relations, which relate the black hole mass $M_{\bullet}$ to a host galaxy property $X$, and a galaxy catalog which contains measurements of the property $X$ for a representative galaxy sample. This estimate is given in Eq.~\ref{eq:BHMF},
where $p(\log M_{\bullet}|X)$ is usually assumed to be a Gaussian with a mean given by a linear relation 
\begin{equation}
    \log \bar{M_{\bullet}} = a_{\bullet} + b_{\bullet} X,
    \label{eq:M-X}
\end{equation}
a standard deviation $\varepsilon_{\bullet}$, and $dn/dX$ is the number density of galaxies per mass proxy $X$.

Since measuring black hole masses is costly and can only be inferred dynamically in a relatively near galaxy sample ($\lesssim 100$~Mpc away), the estimate above can only be achieved by measuring the scaling relation $p(\log M_{\bullet}|X)$ in a smaller local data set, which we will refer to as $\{\vec{d}_i \}$, while $dn/dX$ is measured in a larger galaxy catalog $\{\vec{D}_j \}$, where $i$ and $j$ label each galaxy from a total of $n_{\rm obs}$ and $N_{\rm obs}$ in the respective data set. The larger data set is therefore composed of the observed values $\vec{X}'_j$, which in practice will be the logarithm of the velocity dispersion $\sigma'_{j}$, K-band luminosity $L'_{j}$, the effective radius $R'_{j}$, or functions of these quantities, while $\vec{d}_i = (M'_{\bullet, i}, \vec{X}'_i)$. Note that we distinguish between observed and true values by primed and unprimed variables. For the sake of simplicity in the subsequent discussion, we ignore measurement errors for now and set $\vec{X}' = \vec{X}$, but they are of course reintroduced later on and included in our results.

The task of estimating the GWB from the SMBH mass function can be simplified for two reasons: (i) the fact that the kernel of the mass integral of $h^2_c$ (shown in Eq.~\ref{eq:h2c}) is dominated by a restricted range of $M_{\bullet}$ and $X$ and (ii) that the mass proxy $X$ is always marginalized over in Eq.~\ref{eq:BHMF}. These two features are also suggestive of the origin of any relevant source of error: (i) highlights the narrow range of objects that are of importance for the SGWB prediction, while (ii) elucidates a potential origin for disagreements between different methodologies and mass proxies $X$ --- inconsistencies between the physical meaning of $X$ across data sets. We expand on both of these points below.

To gain intuition regarding (i), consider the $M_{\bullet}-X$ relation of Eq.~\ref{eq:M-X} and the BH mass function from Eq.~\ref{eq:BHMF}. The integrals for the characteristic strain $h^2_c$ or the black hole mass density $\rho_{\rm BH} \equiv \int d\log M_{\bullet}\ M_{\bullet} \frac{dn}{d\log M_{\bullet}}$ are typically of the form
\begin{equation}
\begin{split}
    I(\gamma) =& \int d\log M_{\bullet} \int dX\ M_{\bullet}^{\gamma} \times \\ &\mathcal{N}(\mu = a_{\bullet} + b_{\bullet} X, \sigma = \varepsilon_{\bullet}) \frac{dn}{dX},
\end{split}
\end{equation}
where $\gamma=5/3$ for $h^2_c$ and $\gamma=1$ for $\rho_{\rm BH}$. The integral over the mass can be computed directly, resulting in
\begin{equation}
    I(\gamma) = \int dX \exp\left\{ \frac{1}{2} (\varepsilon_{\bullet} \gamma \ln10)^2 + (a_{\bullet} + b_{\bullet} X)\gamma\ln 10 \right\} \frac{dn}{dX}.
    \label{eq:I_gamma_X}
\end{equation}
The peak $X_{\rm peak}$ of the integrand of Eq.~\ref{eq:I_gamma_X} is given by
\begin{equation}
    \left.\frac{d}{dX} \left(\ln \frac{dn}{dX}\right)\right|_{X=X_{\rm peak}} = -b_{\bullet} \gamma \ln 10.
    \label{eq:Xpeak}
\end{equation}
Note that here $X$ is typically the logarithm of the mass proxy, while the number density of the mass proxy itself ($u\equiv 10^X$) can usually be described as a generalized Schechter function
\begin{equation}
    \phi(u) du = \phi_{*} \left(\frac{u}{u_*}\right)^{\alpha} \frac{e^{-(u/u_*)^{\beta}}}{\Gamma(\alpha/\beta)} \beta \frac{du}{u}.
    \label{eq:sigma_function}
\end{equation}
Changing variables back to $X$ in the number density and substituting in Eq.~\ref{eq:Xpeak} yields
\begin{equation}
    X_{\rm peak} = \frac{1}{\beta \ln10} \ln \left( \frac{\alpha + b_{\bullet} \gamma}{\beta} \right) + X_{*},
\end{equation}
where $X_{*}=\log_{10}u_*$. For example, if $X=\log_{10} \sigma/$km~s$^{-1}$, substituting the best-fit values for the number density from Ref.~\cite{bernardi_MF} ($X_* = 2.2$, $\alpha=0.41$, $\beta=2.59$), and the slope from Ref.~\cite{mcconnell_ma} ($b_{\bullet}=5.64$), results in a peak contribution to the GWB ($\gamma=5/3$) equal to $X_{\rm peak}\sim 2.42$, or $\sigma_{\rm peak}=260~$km s$^{-1}$. This falls in a relatively well measured regime in both the velocity dispersion function and the scaling relation data sets. We also note that the only parameter of the scaling relation that is relevant to $X_{\rm peak}$ is the slope $b_{\bullet}$.

The shape of the integrand suggests a lognormal approximation, obtained by expanding $\ln \left(\frac{dn}{dX}\right)$ to second order around $X_{\rm peak}$. For a Schechter function as in Eq.~\ref{eq:sigma_function}, we find
\begin{equation}
\begin{split}
        \left.\frac{d^2}{dX^2}\left( \ln \frac{dn}{dX} \right) \right|_{X=X_{\rm peak}} = & -(\beta \ln 10)^2 e^{\beta \ln 10 (X_{\rm peak} - X_{*})} \\
        =&-  \beta \ln^2(10) (\alpha + b_{\bullet} \gamma).
\end{split}
\end{equation}
This second order expansion around $X_{\rm peak}$ results in the following analytical expression for the integral of Eq.~\ref{eq:I_gamma_X}
\begin{equation}
\begin{split}
    I(\gamma) =& \sqrt{\frac{2\pi}{-\left.\frac{d^2}{dX^2}\left( \ln \frac{dn}{dX} \right) \right|_{X_{\rm peak}}}} \cdot \left.\frac{dn}{dX}\right|_{X_{\rm peak}}  \times \\ & \exp\left\{ \frac{1}{2} (\varepsilon_{\bullet} \gamma \ln10)^2 + (a_{\bullet} + b_{\bullet} X_{\rm peak})\gamma\ln 10 \right\} \\
    =& \sqrt{\frac{2\pi}{\beta \ln^2(10) (\alpha + b_{\bullet} \gamma)}} \left.\frac{dn}{dX}\right|_{X_{\rm peak}} M_{\bullet,\rm peak}^\gamma e^{\frac{1}{2} (\varepsilon_{\bullet} \gamma \ln10)^2},
\end{split}
\end{equation}
which gives some intuition regarding the galaxies that dominate the integrals and how it depends on the parameters of the mass function and scaling relation.  The square root factor gives a measure of the range of $X$ around the peak that contributes to the integral. One can also see that the result can be quite sensitive to the scatter. 

Turning now to point (ii), Eq.~\ref{eq:BHMF} suggests that consistent changes across both BH $\{\vec{d}_i \}$ and MF $\{\vec{D}_j \}$ catalogs leave the black hole mass function estimate unchanged. Furthermore, given the data sets $\{\vec{d}_i \}$ and $\{\vec{D}_j \}$, using any proxy for the black hole mass $X$ should yield the same estimate of the mass function, as they amount to a straightforward relabeling of the galaxies. The critical assumption, however, is that the quantity $X$ measured in $\{\vec{d}_i \}$ has the same meaning as that measured in $\{\vec{D}_j \}$. Writing this explicitly in Eq.~\ref{eq:BHMF}, 
\begin{equation}
    \frac{dn}{d\log M_{\bullet}} = \int dX \int dx \ p(\log M_{\bullet}|x) p(x|X) \frac{dn}{dX},
    \label{eq:BHMF_diff_x}
\end{equation}
where we have explicitly denoted the mass proxy defined in the BH catalog as $x$ and Eq.~\ref{eq:BHMF} assumed that $ p(x|X) = \delta_D(x-X)$. Suppose now that there is some systematic offset between the mass proxy defined in the BH and MF catalogs, which can be described as a constant shift with a Gaussian scatter. That is 
\begin{equation}
    p(x|X) = \mathcal{N}(\mu=X + \delta, \sigma=\sigma_{\delta}).
\end{equation}

The integral over $x$ can be computed directly. Similarly to Ref.~\cite{Sato-Polito:2023gym} and the integrals discussed above, the shift in the characteristic strain due to the mismatch in mass proxy is given by
\begin{equation}
    \frac{h^2_{c, \delta}}{h^2_{c}} = 10^{\frac{5}{3} b_{\bullet} \delta} e^{\frac{25}{18}\ln^2(10) b_{\bullet}^2\sigma^2_{\delta}}.
    \label{eq:h2c_shift}
\end{equation}
While the ``true'' value of the mass proxy is irrelevant, inconsistencies between the two samples can significantly bias estimates of the GWB amplitude. The impact of a systematic shift in a mass proxy $X$ to the strain prediction is also a strong function of the slope $b_{\bullet}$. Assuming that the slope for $\sigma$, $X_{\rm vir}$, and $L$ or $M_*$ and $b_{\bullet} = 5$, $2.5$, and $1$, respectively, a shift $h^2_{c,\delta}/h^2_c \sim 4$ can be created by an offset $\delta = 0.073$, $0.15$, and $0.36$. In other words, the precision of the strain amplitude determines the required precision of the galaxy property and is a strong function of the slope. Notice that if $\ln(10) b_\bullet \sigma_\delta$ is not small, the scatter in $p(x|X)$ can also have an impact.

\begin{table}[]
    \centering
    \begin{tabular}{|c|c|c|c|} \hline 
        & $\log \sigma$ & $X_{\rm vir}$ & $\log L$ or $M_{*}$\\ \hline
        $b_{\bullet}\sim$ & 5 &  2.5 & 1 \\ \hline 
        $\delta$ & 0.03 & 0.06 & 0.15 \\ \hline 
    \end{tabular}
    \caption{Accuracy requirement on each galaxy property, given current PTA uncertainties on the characteristic strain $h_c$. The quantity $\delta$ is the shift in each property required to produce a change in $h_c$ equal to the width ($\pm \Delta h_c$) of the $90\%$ confidence interval from the NANOGrav 15yr result~\cite{NANOGrav_stoc}. That is, $\frac{\Delta h_c }{h_c} = 10^{\frac{5}{6}b_{\bullet}\delta}$.}
    \label{tab:prop_uncertainty}
\end{table}


\section{Data}\label{sec:data}
To assess the robustness of SGWB estimates and identify potential sources of systematic error in the mass function measurement, we use three primary data sets:
\begin{itemize}
    \item \textbf{Local Sample:} Based on the compilation in Ref.~\cite{R_van_den_Bosch2016}, this sample includes 230 galaxies with black hole mass measurements. Galaxy sizes are primarily reported as $R_{50}$, the major-axis half-light radius measured in elliptical apertures. Velocity dispersions are heterogeneous but generally correspond to the value at the effective radius. We then select only the galaxies for which black hole mass measurements based on stellar or gas kinematics are available, resulting in a final sample of 124. 
    
    In addition to Ref.~\cite{R_van_den_Bosch2016}, we use results from the MASSIVE Survey \cite{2024MNRAS.527..249Q, MASSIVE_2014} as a benchmark for scaling relations in the high-mass regime. The MASSIVE galaxies are selected to be massive early-types based on a K-band magnitude limit from the 2MASS Extended Source Catalog ($M_K < -25.3$ mag), corresponding to stellar masses $M_* \gtrsim 10^{11.5} M_\odot$. The survey obtained deep K-band imaging using WIRCam on CFHT, yielding more accurate size and luminosity estimates than those from the 2MASS catalog. Specifically, they report brighter total magnitudes and larger half-light radii compared to 2MASS values. Size measurements are primarily based on the semimajor axis of the half-light elliptical isophote, although the geometric mean radius is also reported. Velocity dispersions are measured in two ways: central velocity dispersions ($\sigma_c$) using the central fibre of the Mitchell IFS (no attempt is made to correct these dispersions to a common aperture.), and $\sigma_e$ values as luminosity-weighted averages within $R_e$, based on radii corrected to agree with the NSA catalog. 

    \item \textbf{6dFGSv Fundamental Plane (FP) Sample:} A subset of the 6dF Galaxy Survey containing approximately $10^4$ early-type galaxies used for Fundamental Plane studies (see Refs.~\cite{6dFGSv_FP, 6dFGSv_data}). The effective radii are derived using GALFIT applied to 2MASS images to obtain PSF-corrected 2D S\'ersic fits. Galaxies are selected based on spectral fitting favoring early-type templates within the 6dF fibre aperture, followed by visual inspection to remove objects with significant disc contamination. Central velocity dispersions are measured using a Fourier cross-correlation method and corrected to a common aperture of $R_e/8$ using a power-law relation from Ref.~\cite{1995MNRAS.276.1341J}, with a lower cut of $\sigma_0 \geq 112$ km/s imposed due to instrumental resolution, magnitude limit of $m_K\leq 12.55$, and $z<0.055$.

    \item \textbf{6dFGS DR3 Catalog:} The full DR3 catalog (125,071 galaxies) is used to estimate black hole, velocity, and luminosity functions based on scaling relations. For this sample, galaxy sizes and magnitudes come directly from the 2MASS Extended Source Catalog. The 6dFGS catalog is complete to total extrapolated K-band magnitude $m_K\leq 12.65$, with a median redshift of $z_{1/2}=0.053$ and covering $41\%$ of the sky.
\end{itemize}

Throughout this work, we refer to the galaxy sample with black hole mass measurements as the ``BH'' catalog, obtained from the compilation in Ref.~\cite{R_van_den_Bosch2016}, while both the BH and MASSIVE samples are referred to as local samples. The 6dFGSv fundamental plane catalog is labeled ``FP'' and the full 6dFGS from which the mass function is measured is the ``MF'' catalog. 

\begin{figure*}[t]
    \centering
    \includegraphics[width=0.9\linewidth]{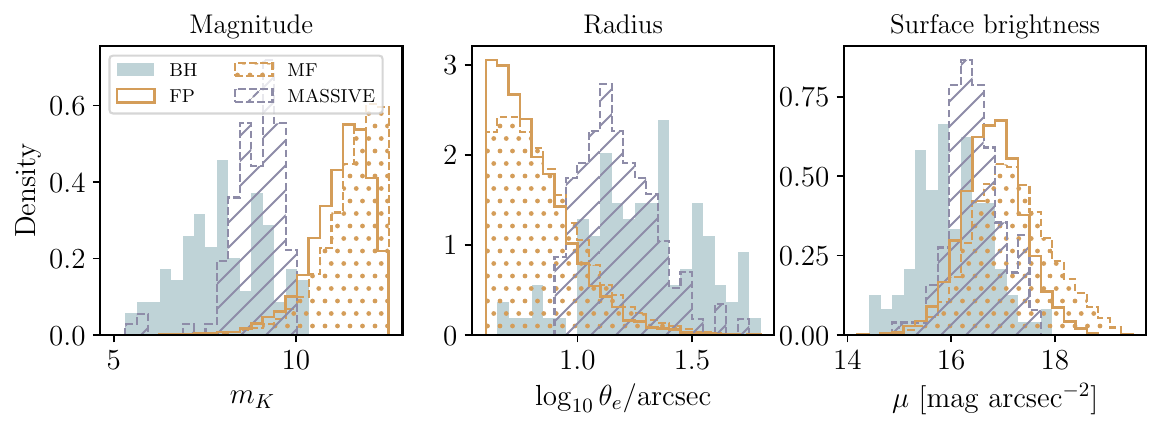}
    \caption{Histograms of observed galaxy properties across the BH~\cite{R_van_den_Bosch2016}, MASSIVE~\cite{2024MNRAS.527..249Q}, FP~\cite{6dFGSv_data}, and MF~\cite{6dFGS_DR3} catalogs for the objects used in this analysis.}
    \label{fig:obs_props}
\end{figure*}

In order to standardize the samples as much as possible and avoid the impact of systematic biases, we apply consistent cuts on the FP and MF catalogs. We select only galaxies with $z<0.055$ in the MF catalog, resulting in a reduction to about half the sample, which is still sufficiently large for a robust mass function measurement. We also require $m_K<12.55$, and $\log \theta_e/$arcsec$>0.6$, such that the radius is not affected by the PSF (which has a FWHM of $\sim 3$~arcsec), and $\log \sigma/$km s$^{-1}>2.05$ across both local and FP catalogs due to the resolution limit of the FP sample. The distribution of observed properties once these cuts are applied are shown in Fig.~\ref{fig:obs_props}. Finally, we adopt Ref.~\cite{kcorrection_2010} for K-correction.

One useful feature of this choice of data set is the redundancy between them, which is summarized in Table~\ref{tab:properties}. For instance, galaxies in the local, FP, and MF samples have all been measured in 2MASS. Both Refs.~\cite{R_van_den_Bosch2016} and \cite{2024MNRAS.527..249Q} have analyzed independent sets of nearby galaxies. Based on the discussion in Sec.~\ref{sec:general}, a consistent under- or over-prediction of a given galaxy property across catalogs would not bias the stochastic background prediction. However, systematic biases across the local and full samples are known to exist, which we discuss below.

\begin{table}[h]
    \centering
    \begin{tabular}{|c|c|c|} \hline 
        Local galaxies & FP catalog & MF catalog \\ \hline 
        $M_{\bullet}^{(1)}$ & --- & --- \\ \hline
        $\sigma^{(1)}$, $\sigma^{(2)}$ & $\sigma^{(3)}$ & --- \\ \hline
        $L^{(1)},L^{(2)}, L^{(4)}$ & $L^{(3)}\equiv L^{(4)}$ & $L^{(4)}$ \\ \hline
        $R^{(1)},R^{(2)}, R^{(4)}$& $R^{(3)}, R^{(4)}$ & $R^{(4)}$ \\ \hline
    \end{tabular}
    \caption{(1) BH sample~\cite{R_van_den_Bosch2016}, (2) MASSIVE sample~\cite{2024MNRAS.527..249Q}, (3) 6dFGSv~\cite{6dFGSv_data, 6dFGSv_FP}, (4) 2MASS~\cite{2006AJ....131.1163S, 6dFGS_DR3, 6dFGS_DR1}. Note that the full catalog corresponds to the 6dFGS sample, which include the 2MASS properties.}
    \label{tab:properties}
\end{table}

\subsection{Corrections to luminosity and radius}\label{subsec:LR_corr}
Refs.~\cite{R_van_den_Bosch2016} and \cite{2024MNRAS.527..249Q} compared their inferred values of luminosity and radius with those inferred by 2MASS in local galaxy samples, which can shed light on potential systematic errors. In particular, it has been noted that 2MASS appears to systematically underestimate the total luminosity and effective radius \cite{2007ApJ...662..808L, Schombert_Smith_2012}. As previously noted, a constant underestimate would not bias the SMBH mass function estimate. However, \cite{R_van_den_Bosch2016} and \cite{2024MNRAS.527..249Q} find that the correction does have some dependence on apparent magnitude and angular size. We therefore correct the magnitude and size in 2MASS in order to match the properties measured in both references and compare the 3 properties to study this effect.

The relation between the 2MASS magnitude $m_K$ and angular effective radius $\theta_e$ and those found in Refs.~\cite{R_van_den_Bosch2016} and the MASSIVE sample are described by a linear fit
\begin{equation}
    X - X_{\rm pivot} = a_{X}+ b_{X}(X^{\rm 2MASS} - X_{\rm pivot}),
\end{equation}
where $X=m_K, \log(\theta_{e}/\text{arcsec})$ and the chosen pivot values are $X_{\rm pivot}=9, 1.2$. The best-fit parameters and uncertainties are given in Table~\ref{tab:calibration_fits}.

\begin{table}[h]
    \centering
    \begin{tabular}{|c|c|c|c|c|}
    \hline
     & \multicolumn{2}{c|}{BH} & \multicolumn{2}{c|}{MASSIVE} \\ \hline
          & $m_K$ & $\log\theta_e$ & $m_K$  & $\log\theta_e$ \\ \hline
        $a$ & $-0.26\pm0.03$ & $0.17\pm0.01$ & $-0.292\pm0.009$  & $0.072\pm0.006$\\ \hline
        $b$ & $1.02\pm0.02$ & $1.13\pm0.04$ & $0.950\pm0.017$  & $0.96\pm0.04$\\ \hline
    \end{tabular}
    \caption{Parameters of the linear fit used to correct the 2MASS magnitude and radius based on the same parameters measured in Refs.~\cite{R_van_den_Bosch2016} and \cite{2024MNRAS.527..249Q}.}
    \label{tab:calibration_fits}
\end{table}

We fit the BH parameters from the fiducial subset of galaxies used in this work, while the MASSIVE values are taken directly from Ref.~\cite{2024MNRAS.527..249Q}. Note that the typical magnitudes and radii of galaxies in the local and 6dFGS catalogs are in very different regimes, as shown in Fig.~\ref{fig:obs_props}. Using the corrections above result in extrapolating the fit to significantly fainter magnitudes and smaller radii, far beyond measured values. While there is a consensus that the 2MASS K-band magnitudes are fainter by roughly $\sim 0.3$~mag (see also, e.g., \cite{Schombert_Smith_2012}), the SGWB estimate is very sensitive to the inferred slope of this relation. There is a much larger disagreement on the inferred angular size of the galaxy, indicating a less robust measurement. The two references, \cite{R_van_den_Bosch2016} and \cite{2024MNRAS.527..249Q}, find significantly different corrections, with slopes $>1$ and $<1$, respectively. We take both of these options into account and consider them illustrative of the inherent uncertainties in total luminosity and effective radius measurements, but stress the arbitrariness of the results inferred from extrapolating these linear relations beyond the local sample, as well as the importance of more careful studies quantifying these biases. 

\subsection{Corrections to velocity dispersion}\label{sec:vdisp_corr}
The velocity dispersion in the larger catalog in which the mass functions are measured (e.g., SDSS, 6dFGSv) are corrected to an aperture size of $\theta_e/8$, while in the local sample the velocity dispersion is typically reported at $\theta_e$. Ref.~\cite{6dFGSv_data} uses the empirical correction
\begin{equation}
    \frac{\sigma(\theta)}{\sigma_{\rm ap}} = \left(\frac{\theta}{\theta_{\rm ap}}\right)^{\gamma_{\rm ap}},
\end{equation}
where $\theta_{\rm ap} = 3.35$~arcsec is the radius of the fibre, $\sigma_{\rm ap}$ is the measured velocity dispersion at the aperture radius, and $\sigma(\theta)$ is the value corrected to an angular radius $\theta$, which in this case is $\theta=\theta_e/8$, and $\gamma_{\rm ap}$ is the slope of the inner velocity dispersion profile. The correction used in \cite{6dFGSv_data} is $\gamma_{\rm ap} = -0.04$, derived from Ref.~\cite{1995MNRAS.276.1341J}, and is commonly adopted in the literature. There are a variety of more modern studies measuring the velocity dispersion profiles from different galaxy samples (see, e.g., \cite{2006MNRAS.366.1126C, 2017A&A...597A..48F, 2018MNRAS.473.5446V, 2023RAA....23h5001Z}). A robust conclusion found by such studies is that the inner profile rises towards the center for the massive/bright galaxies that are relevant to the GWB estimate, with average slopes found to be $-0.07 \leq \gamma_{\rm ap} \leq -0.03$. This implies an average shift in the velocity dispersions reported in the full catalog of $\delta=-\gamma_{\rm ap} \log8$. Using Eq.~\ref{eq:h2c_shift}, the observed range of $\gamma_{\rm ap}$ corresponds to a reduction in the inferred strain of $0.5 \leq \frac{h_{c,\delta}}{h_c} \leq 0.75$, and $\frac{h_{c,\delta}}{h_c} = 0.68$ for the fiducial $\gamma_{\rm ap} = -0.04$.

\section{Measuring the black hole mass function}
\subsection{Fitting conditional probabilities}
We assume that the intrinsic relation between any two properties $X$ and $Y$ can be described by a normal distribution
\begin{equation}
    p(Y|X, \vec{\eta}) = \frac{1}{\sqrt{2\pi} \varepsilon} \exp\left[-\frac{( Y - \bar{Y}(X;a, b))^2}{2\varepsilon^2} \right],
\end{equation}
where the mean is given by $ \bar{Y} (X; a, b) = a + b X$, and $\vec{\eta}= (a, b, \varepsilon)$ are the model parameters that describe the scaling relation. 
Note that $X$ and $Y$ are typically the logarithm of one of the observed galaxy properties ($\sigma$, $L$, or $X_{\rm vir}$), so the distribution of the galaxy property is assumed to be lognormal.

For each $i$-th object in the local galaxy sample, we observe $\vec{d}_i=(X'_i, Y'_i)$. Assuming a Gaussian likelihood for the measurement, the probability of making the full set of observations is given by
\begin{equation}
\begin{split}
    p(\{\vec{d}_i \} | \vec{\eta}) =& \prod^{n_{\rm obs}}_{i=1} p(X'_i, Y'_i|\vec{\eta}) \\ =& \prod^{n_{\rm obs}}_{i=1} \int dX_i dY_i\ p(X'_i, Y'_i|X_i, Y_i) p(Y_i|X_i, \vec{\eta}) p(X_i| \vec{\eta})\\ =& \prod^{n_{\rm obs}}_{i=1} \frac{1}{\sqrt{2\pi} \sigma_{{\rm tot},i}} \exp\left[-\frac{(\log Y'_i - \log\bar{Y}(X') )^2}{2\sigma^2_{{\rm tot},i}}\right],
    \label{eq:likelihood}
\end{split}
\end{equation}
where a diagonal covariance $\Sigma=(\sigma^2_{X,i}, \sigma_{Y,i}^2)$ is assumed for $p(X'_i, Y'_i|X_i, Y_i)$ in the third equality. We also defined $\sigma^2_{{\rm tot},i} = \sigma_{Y,i}^2 + b^2 \sigma_{X,i}^2 +\varepsilon^2$, and assumed that $p(X_i| \vec{\eta})$ is uniform. Note that this expression is straightforward to generalize in case we wish to consider, for instance, correlated measurement errors, which is often present for $L$ and $R$ measurements. In this case, 
\[
\Sigma = \begin{pmatrix}
\sigma_{X_i}^2 & \rho \sigma_{X_i} \sigma_{Y_i} \\
\rho \sigma_{X_i} \sigma_{Y_i} & \sigma_{Y_i}^2
\end{pmatrix}.
\]

The observed catalog will typically contain a biased selection of galaxies. In the presence of selection effects, the likelihood in Eq.~\ref{eq:likelihood} must be normalized by the fraction of objects that would be observed given set of parameters $\vec{\eta}$ (see Ref.~\cite{Mandel:2018mve}). Given a known selection function $f_{\rm det} (X',Y')$, the probability of detecting a galaxy with true parameters $X,Y$ is
\begin{equation}
    p_{\rm det} (X,Y) = \int p(X',Y'|X,Y) f_{\rm det} (X',Y') dX' dY'.
\end{equation}
We then divide the likelihood in Eq.~\ref{eq:likelihood} by the normalization
\begin{equation}
    \alpha(\vec{\eta}) = \int dX dY p_{\rm det} (X,Y) p(X,Y|\vec{\eta}).
\end{equation}
We will typically assume that the selection criteria between $X$ and $Y$ are independent and the measurements are uncorrelated, so that $p_{\rm det}(X,Y) = p_{\rm det}(X) p_{\rm det}(Y)$. Consider a selection effect only on $Y$. A relevant example is the presence of some threshold value $Y_{\rm th}$ for the observed $Y'$, above which the galaxy is always included in the catalog. In this case, $f_{\rm det}(Y')=\Theta(Y'-Y_{\rm th})$, which describes the selection effect for velocity dispersions in the fundamental plane sample. 

One important caveat is that, while these selection effects are well characterized in the larger galaxy catalogs, they are less known in the local galaxy sample. We tentatively assume that there are no selection effects in the local sample. We also note that, in practice, the selection effect in the $X$ component is only relevant in the regime where its measurement error is comparable to the scatter in the $Y$ direction, that is, $b^2\sigma_X^2 \sim \sigma_Y^2 + \varepsilon^2$. However, this is typically not the case for the cases considered in this work, which are dominated by intrinsic scatter. It is therefore a reasonable approximation to neglect selection effects on $X$. 

In practice, the selection effects present in the FP and MF samples are the ones described in Sec.~\ref{sec:data}: $\log \sigma>\log \sigma_{\rm th} = 2.05$, $m_K<m_{K,{\rm lim}}=12.55$, and $\log \theta_e>\log \theta_{e,{\rm lim}}=0.6$. Assuming that the errors in the velocity dispersion measurement are uncorrelated with the other properties, including this selection effect is straightforward, since the detection probability can be directly computed:
\begin{equation}
    p_{\rm det} (\sigma) = \frac{1}{2} \left[1 - \text{erf} \left(\frac{\sigma_{\rm th} - \sigma}{\sigma_{\sigma}} \right)\right].
\end{equation}

A magnitude limit implies that, for each absolute magnitude or luminosity, there is a maximum distance $d_{\rm max}(L)$ out to which a galaxy could be observed. That is,
\begin{equation}
    \log\frac{d_{\rm max}}{\rm pc} (L) = \frac{1}{2} \log\left( \frac{L}{L_{K, \odot}} \right) +  \frac{1}{5}\left(m_{K, {\rm lim}} - M_{K, \odot}\right) +1,
\end{equation}
and the redshift corresponding to $d_{\rm max}$ is $z_{\rm max}(L)$. If the survey is complete out to some redshift $z_{\rm lim}$ and the galaxies have uniform number density, then
\begin{equation}
    p_{\rm det}(L) = 
    \begin{cases} \frac{V(z_{\rm max}(L))}{V(z_{\rm lim})}, \quad \text{for } z_{\rm max}<z_{\rm lim} \\
    1, \quad \text{otherwise.} \end{cases}
    \label{eq:pdet_vol}
    \end{equation}
Similarly, if we impose a minimum angular size $\theta_{e, {\rm lim}}$ for the galaxies, this implies that for each radius there is a maximum distance out to which it could be observed
\begin{equation}
    d_{A,{\rm max}}(R) = \frac{R}{\theta_{e, {\rm lim}}},
\end{equation}
and $p_{\rm det}(R)$ is defined as in Eq.~\ref{eq:pdet_vol}.

\subsection{Mass function}\label{subsec:MF}
We consider 3 main galaxy properties $X$ as proxies for the BH mass: velocity dispersion $\sigma$, K-band luminosity $L$, and virial parameter $X_{\rm vir} \equiv \log L + \alpha_{\rm vir} \log R$. We choose to define $X_{\rm vir}$ such that it is consistent with fundamental plane measurements, i.e., $\log \sigma \propto X_{\rm vir}$. Hence, we adopt the empirical value $\alpha_{\rm vir} = -0.8$~\cite{6dFGSv_FP, 2024MNRAS.527..249Q}, but discuss the impact of such a choice in Sec.~\ref{subsec:sig_L_R}

The black hole mass function is estimated from the data set discussed in Sec.~\ref{sec:data} through  an end-to-end bootstrap procedure with $N_{\rm bs}=10^4$ iterations. In each iteration:
\begin{enumerate}
    \item The local sample and full 6dFGS catalog are each resampled with replacement. When the velocity dispersion is used, the fundamental plane catalog is also resampled with replacement.
    \item Define mass proxy in local and full catalogs. Since the full 6dFGS does not have velocity dispersion, if that is used as the mass proxy, we first fit the relation between $X_{\rm vir}$ and $\sigma$ in the FP sample. For each galaxy in the resampled 6dFGS catalog, we then assign a velocity dispersion $\hat{\sigma}(X_{{\rm vir}, i})$ based on its virial parameter using the FP fit.
    \item Fit the black hole scaling relation $M_{\bullet}-X$ in the local sample.
    \item For each galaxy in the resampled 6dFGS catalog, a black hole mass $\bar{M}_{\bullet, i}$ is assigned based on the scaling relation fit in item 3.
    \item A Gaussian random component $n_i$ is added to the inferred mass $M_{\bullet,i}=\bar{M}_{\bullet, i} + n_i$ value (and $\hat{\sigma}_i$, if used) to reproduce the intrinsic scatter found in the scaling relation. The additional noise term is calibrated such that, when combined with measurement uncertainties, the total dispersion matches the intrinsic scatter, resulting in a practical correction for Eddington bias. That is, if $X'$ is the measured value of $X$ with error $\sigma_X$, and the scaling relation is $Y\sim \mathcal{N}(\mu=a+bX, \sigma=\varepsilon_Y)$, then the added noise is sampled from $n_i\sim \mathcal{N}\left(\mu=0, \sigma=\sqrt{\varepsilon_Y^2 -b^2\sigma^2_{X,i}}\right)$.
    \item The black hole mass function (and velocity function) are computed using the 1/$V_\text{max}$ method. For each galaxy, the maximum volume $V_\text{max}$ out to which it could be observed is computed using its K-band magnitude, given that the survey limit is of $K < 12.55$ and choose a minimum effective radius  $\log_{10} \theta_e/{\rm arcsec} > 0.6$. The minimum radius was imposed to reduce the effect the smoothing by the 2MASS beam. 
\end{enumerate}
The results are stored at each iteration and used to compute the median, 16th, and 84th percentiles of the velocity and mass functions in each bin.


\section{Results}
\subsection{Uncertainties on galaxy properties and impact on the SGWB}

\begin{figure}
    \centering
    \includegraphics[width=0.9\linewidth]{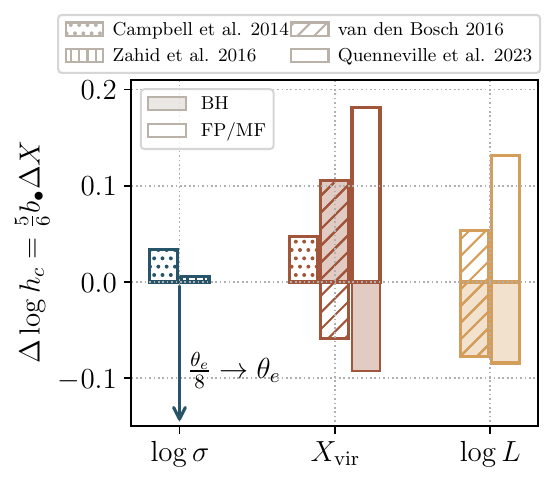}
    \caption{Variation in the characteristic strain value from shifts $\Delta X$ in the mass proxy $X=(\log\sigma, X_{\rm vir}, \log L)$ found in the literature, shown in dark blue, red, and yellow respectively. Here $\Delta X$ is the magnitude of the offsets found between different data sets in the corresponding reference and the y-axis shows the log of the ratio between the inferred characteristic strains. Unfilled bars correspond to corrections applied to the FP or MF sample, while filled are the corrections to the BH sample. The arrow shows the shift induced by correcting the velocity dispersion to the value at the effective radius $\theta_e$ for  $\gamma_{\rm ap} = -0.04$, discussed in Sec.~\ref{sec:vdisp_corr}.}
    \label{fig:shift_hc}
\end{figure}

\begin{figure*}[t]
    \centering
    \includegraphics[width=0.65\linewidth]{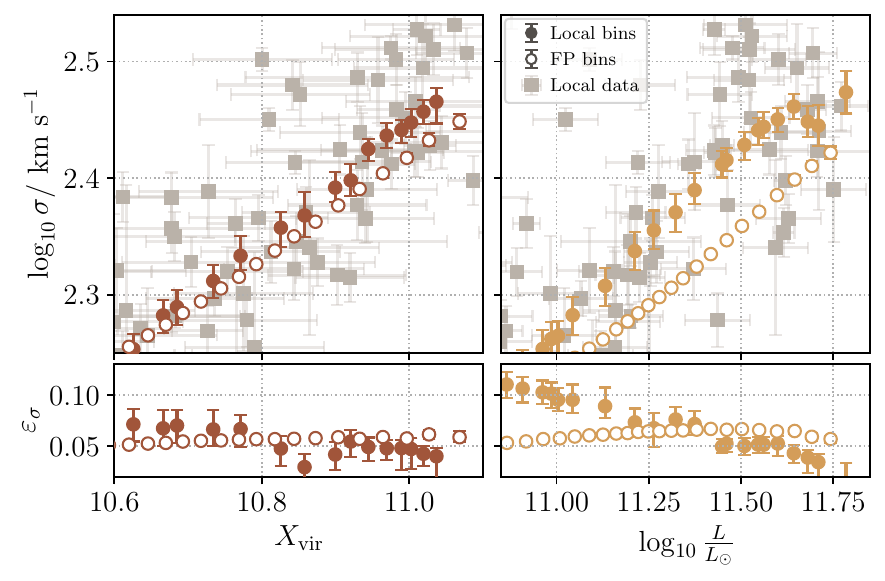}
    \caption{Relation between virial parameter $X_{\rm vir}$ and velocity dispersion $\sigma$ on the left and luminosity $L$ and $\sigma$ on the right. The bottom panel shows the inferred intrinsic scatter. The filled and unfilled circles correspond to the measured amplitude in bins for the BH and FP catalogs, while the background gray squares are the data points for the BH sample. In this figure $L$ and $R$ are the uncorrected values from the 2MASS catalog.}
    \label{fig:sig_LR_L}
\end{figure*}

To our knowledge, there are no studies quantifying systematic shifts in the galaxy properties across the relevant regimes that capture galaxies in both the BH and MF catalogs (i.e., quantifying $p(x|X)$). Nevertheless, to get a sense for the magnitude of potential errors, we compare the impact on the characteristic strain of shifts in galaxy properties found across a few references in the literature in Fig.~\ref{fig:shift_hc}. The equivalent value for the mass density is given by
\begin{equation}
    \Delta \log \rho_{\rm BH} = b_\bullet \Delta X = 6/5\ \Delta \log h_c.
\end{equation}

The dotted bars correspond to the results found in Ref.~\cite{6dFGSv_data}, where the $\sigma$ case shows a comparison with the velocity dispersions in SDSS~\cite{SDSS_vdisp_DR8} (found in Table 3 of \cite{6dFGSv_data}, along with comparisons with other values in the literature) and the $X_{\rm vir}$ case shows a comparison with \cite{pahre_1999}\footnote{Note that Ref.~\cite{6dFGSv_data} uses a slightly different quantity than $X_{\rm vir}$. Ref.~\cite{6dFGSv_data} defined $X_{\rm FP} =  \log R -A\avg{\mu_e}$ with $A=0.3$. In this case, $X_{\rm FP} = 2.5 A\log L + (1-5A)\log R$, which gives $X_{\rm FP} =0.75 \log L -0.5\log R$ for $A=0.3$. We therefore compute the shift using $X_{\rm vir} = X_{\rm FP}/0.75$.} (Fig.~12 of \cite{6dFGSv_data}). The vertical hatching corresponds to the comparison presented in Ref.~\cite{zahid_2016} between SDSS and the Smithsonian Hectospec Lensing Survey (SHELS). In general, comparisons between different estimates of velocity dispersion across galaxy surveys that extend beyond the local (BH) sample find consistent results within the uncertainties required to estimate the SGWB.

Evidence for larger shifts in $L$ and $R$ are typically found. The filled diagonal hatching and unhatched  show the shift induced by the corrections to apparent magnitudes and angular radii discussed in Sec.~\ref{subsec:LR_corr}. Filled and unfilled bars correspond to shifts due to corrections applied to the BH and FP/MF samples respectively. The precise value of the shift is computed as the mean of the difference between the 2MASS and corrected luminosity and physical radius for each object, after selecting galaxies with $\log \sigma/$km s$^{-1}>2.3$. The final shift in the strain prediction relative to the uncorrected 2MASS value is the sum of the between the filled and unfilled corrections. While the total magnitude of the corrections are $\sim 0.08-0.13$, since the corrections to the BH and FP/MF partially cancel, this results in a final shift relative to the uncorrected 2MASS of only about $\delta\sim 0.03-0.04$, leading to the small change in the final result of Fig.~\ref{fig:hc_summary}. Note that \cite{R_van_den_Bosch2016} and \cite{2024MNRAS.527..249Q} predict very different corrections to both catalogs that end up somewhat canceling out. This is especially noticeable for $X_{\rm vir}$ where the corrections (\cite{R_van_den_Bosch2016} and \cite{2024MNRAS.527..249Q}) even have opposite signs. Given the caveats discussed in Sec.~\ref{subsec:LR_corr} and the fact that these corrections are both substantial and poorly measured, the consistency between these estimates appears to be a coincidence. 

Finally, the average shifts in the luminosity are derived in a similar manner as $X_{\rm vir}$. Fig.~\ref{fig:shift_hc} indicates a slightly smaller shift in the strain due to magnitude corrections compared to $X_{\rm vir}$, owing to the shallower slope of $M_{\bullet} - L$. We will discover however that selection effects are much more dramatic for the luminosity and that those uncertainties will dominate over the small advantage apparent here.

\subsection{$\sigma$, $L$, and $R$ in local and FP catalogs}\label{subsec:sig_L_R}

A straightforward consistency check that the local and distant samples should satisfy is that the empirical scaling relations between galaxy properties must be the same. This should be satisfied if the following conditions are true:
\begin{enumerate}
    \item the galaxy properties are consistently defined between local and distant catalogs and
    \item the galaxy samples are consistent. Either because:
    \begin{enumerate}
         \item[a)] all galaxies satisfy the same scaling relation or
         \item[b)] local and distant samples contain the same galaxy populations
    \end{enumerate}
\end{enumerate}

In Fig.~\ref{fig:sig_LR_L}, we compare the relation between $\sigma$, the virial parameter $X_{\rm vir}$, and $L$ in the local and fundamental plane samples. For clarity, we only show the case for one choice of galaxy property definition --- the uncorrected 2MASS magnitude and radius and $\sigma$ at $\theta_e/8$ --- but we comment on the impact of the corrections and include all options in the final results in Sec.~\ref{sec:MF}. The values for the colored points are obtained by fitting the scaling relation in bins and the median and 1$\sigma$ confidence intervals are obtained from bootstraping (as in Sec.~\ref{subsec:MF}). Note that the bin width is $5\times$ the bin spacing, so the points are highly correlated. 

None of the galaxy properties (left or right panel) perfectly satisfy this consistency check, particularly in the high velocity dispersion regime, which is the most consequential for estimates of the GWB. We re-iterate that the relevant point in Fig.~\ref{fig:sig_LR_L} is the difference between filled and unfilled dots, since this is the origin of the inconsistency between estimates of the SMBH mass function based on different mass proxies. The magnitude and radius corrections discussed in Sec.~\ref{sec:data} tend to exacerbate the gap between local and FP scaling relations, therefore leading to a larger difference between inferences of the mass function based on different mass proxies. 

\begin{figure}
    \centering
    \includegraphics[width=0.9\linewidth]{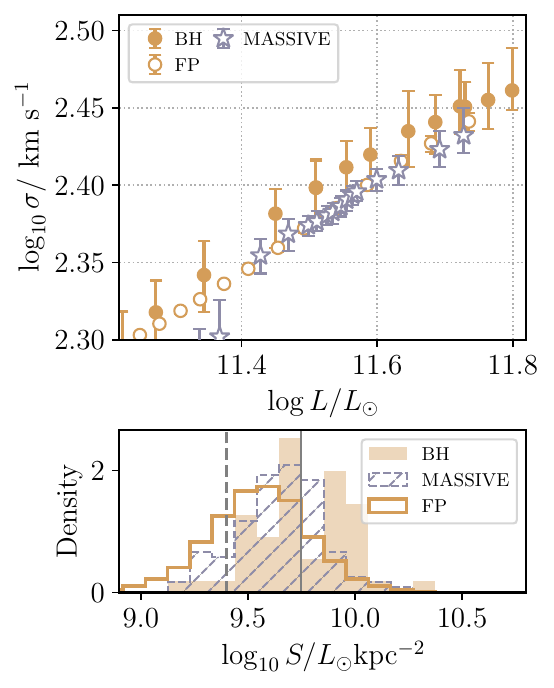}
    \caption{Effect of differences in surface brightness across galaxy samples. The distribution of surface brightnesses in the black hole~\cite{R_van_den_Bosch2016}, MASSIVE~\cite{2024MNRAS.527..249Q}, and fundamental plane~\cite{6dFGSv_FP} samples are shown in the bottom panel. The vertical solid line shows the upper limit to the surface brightness applied to the BH sample ($\log_{10}S/L_{\odot}$kpc$^{-2}<9.75$) and the vertical dashed shows the lower limit applied to the FP sample ($\log_{10}S/L_{\odot}$kpc$^{-2}>9.45$). The top panel shows the relation between velocity dispersion $\sigma$ and K-band luminosity $L$ in local and fundamental plane samples once the aforementioned surface brightness cuts are applied, while the MASSIVE sample is used with no cuts.}
    \label{fig:sigma_L_SB}
\end{figure}

\begin{figure}
    \centering
    \includegraphics[width=1\linewidth]{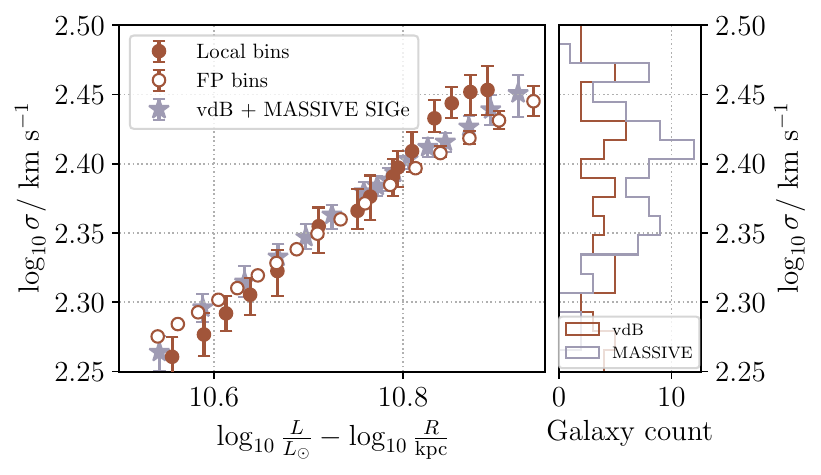}
    \caption{Relation between velocity dispersion $\sigma$ and virial combination $L/R$ in the local BH sample (filled red circle), FP sample (unfilled red circle), and the local BH sample combined with the MASSIVE sample (filled stars). The panel on the right shows the number of galaxies in each $\sigma$ bin, showing the large increase in sample size from the MASSIVE survey.}
    \label{fig:sigma_LR_MASSIVE}
\end{figure}

The right panel of Fig.~\ref{fig:sig_LR_L} shows a large inconsistency in the relation between velocity dispersion and K-band luminosity between the local and FP catalogs. This inconsistency is clear both as an overall offset in the $\sigma$ for a given $L$ and in the scatter, and is not significantly affected by the different magnitude corrections. This is the origin of the discrepancy between predictions of the SMBH mass function from scaling relations based on $\sigma$ or $L$ (see, e.g., Ref.~\cite{tundo2007}). In spite of these large disagreements, once the virial combination is computed, local and FP catalogs are much more consistent, shown in the panel on the left, although a disagreement in the predicted SMBH mass function is still expected due to the difference in $X_{\rm vir}$ vs $\sigma$ at high $X_{\rm vir}$. 

Given the virial theorem $\sigma^2 = \frac{GM}{R}$, the assumption of an approximately constant mass-to-light ratio $M/L$ leads to the expectation of $\sigma \propto (L/R)^{1/2}$. A constant surface brightness $S = L/4\pi R^2$ then leads to the Faber–Jackson relation $\sigma \propto (S\cdot L)^{1/4}$. The inconsistency between $\sigma-L$ across the two catalogs, while the virial relation is approximately satisfied, may then be due to the samples having a different average surface brightness. 

To explore this possibility, we begin by applying cuts to the BH and FP samples based on the surface brightness of galaxies. The surface brightness is computed as $S=L/\pi R^2k_{ba}$, where $k_{ba}$ is the minor to major axis ratio provided in the 2MASS catalog. The distribution of $S$ in each sample is shown in the bottom panel of Fig.~\ref{fig:sigma_L_SB}, and it is clear that the FP, MASSIVE, and BH samples have ascending median surface brightness. Galaxies in the FP sample are selected such that $\log_{10} S/L_{\odot}$kpc$^{-2}>9.45$ and $\log_{10}S/L_{\odot}$kpc$^{-2}<9.75$ in the BH sample, such that the values roughly match the MASSIVE sample, most of the difference between $\sigma-L$ disappears, shown in the upper panel. 

It is thus clear that the difference in surface brightness is the source of most of the difference in $\sigma-L$ between the local and FP samples. ``Accounting for sample differences in surface brightness'' is equivalent to accounting for both the luminosity and the radius simultaneously, and hence we recover the fundamental plane scenario $\sigma-X_{\rm vir}$. The scaling relations $\sigma-X_{\rm vir}$ and $\sigma-L$ correspond to projections of different rotations of the fundamental plane, where the former is the edge-on view. The inferred intrinsic scatter of the $\sigma-L$ relation therefore also captures the range of surface brightnesses of each sample; for a fixed luminosity, a higher surface brightness corresponds to a smaller radius and, from the fundamental plane relation, a higher $\sigma$. We discuss this in further detail in appendix~\ref{app:FP}.

The remaining difference in $\sigma-X_{\rm vir}$ between the BH and FP samples is a consequence of the samples having different measured fundamental planes. Fitting each sample yields $\alpha^{\rm BH}_{\rm vir} = -0.95 \pm 0.006$ and $\alpha^{\rm FP}_{\rm vir} = -0.755 \pm 0.001$, both with similar intrinsic scatter of $\varepsilon_{\sigma} = 0.06$. It is therefore not possible to view both fundamental planes edge-on with the same projection. Given the magnitude of the systematic shifts $\delta$ discussed in Sec.~\ref{sec:data}, this is hardly surprising. While this is likely an important source of the remaining inconsistencies, it is also possible that the samples are intrinsically different.

As a final check, we compare the $\sigma-X_{\rm vir}$ relations when the MASSIVE sample is included in the fit. With a volume-limited sample of the most massive galaxies in the local universe, adding it to the BH sample should have an impact if the source of the discrepancy is due to sample selection. In Fig.~\ref{fig:sigma_LR_MASSIVE}, we show the $\sigma-X_{\rm vir}$ relation in the local sample when the BH sample is combined with the MASSIVE catalog, and compare it to the results from the left panel of Fig.~\ref{fig:sig_LR_L}. While Ref.~\cite{2024MNRAS.527..249Q} reports velocity dispersions, luminosities, and radii, uncertainties for these quantities are not publicly available. We therefore assign uncertainties by sampling a normal distribution with the same mean and variance as the galaxies in the sample from Ref.~\cite{R_van_den_Bosch2016}. There is a clear shift in the $\sigma-X_{\rm vir}$ relation when the MASSIVE galaxies are included which suggests that selection differences between the BH and MF catalogs also play a role.

To get an approximate sense for the relative importance between systematic shifts in the mass proxy and selection effects, we can compare the corrections to $X_{\rm vir}$ and the scatter of the scaling relation. If we suppose that typical systematic corrections will lead to shifts $\delta\sim0.05-0.1$, then the corresponding shift in the strain is $\Delta \log h_c \sim 0.1-0.2$. To estimate the impact of potential selection effects, we follow a similar discussion as in App.~\ref{app:FP}. One could argue that $\sigma$ depends on some other property $Y$ in addition to $X_{\rm vir}$, and that the distribution of $Y$ differ in BH and FP/MF samples. The inferred scatter $\varepsilon_{\sigma}=0.06$ would therefore be absorbing the range of $Y$ in the samples. The average value of $Y$ may be arbitrarily different between the two samples, and therefore we cannot estimate this effect from first principles. However, assuming that the MF sample has a known selection and the BH sample is a biased subset of this distribution, then a natural magnitude for the shift from selection effects is of the order of the scatter. If we suppose that half of the inferred $\varepsilon_{\sigma}$ comes from the width of the $Y$ distribution, then the largest shift is $\delta \sim 0.04$, leading to $\Delta\log h_c=0.08$. Hence, while it appears that in this scenario observational systematics may be more important, selection effects can be comparable in magnitude. 

In summary, we find that, while there is evidence that systematic shifts in the mass proxy are significant, selection effects may also be playing a role.

\subsection{Velocity Dispersion Function (VDF)}\label{sec:VDF}
\begin{figure}[t]
    \centering
    \includegraphics[width=0.9\linewidth]{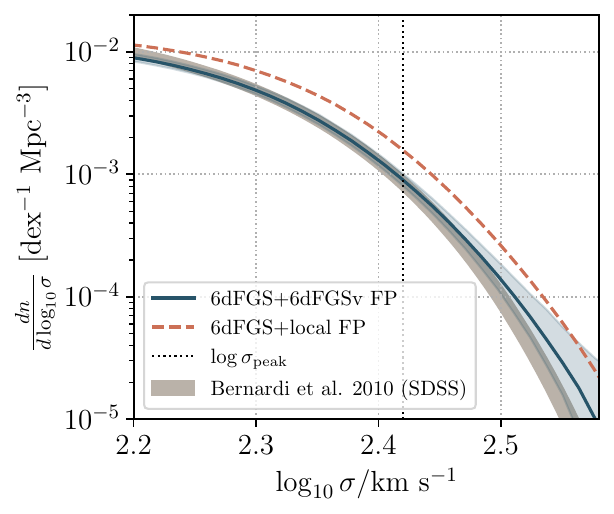}
    \caption{Velocity dispersion function from Ref.~\cite{bernardi_MF} measured from SDSS (gray band) and from the 6dFGS (blue solid line), measured in this work. The orange dashed line shows the VDF when the local relation $\sigma-X_{\rm vir}$ is used, illustrating the same discrepancy shown in Figs.~\ref{fig:BHMF_hc} and \ref{fig:hc_summary}.}
    \label{fig:VDF}
\end{figure}

\begin{figure*}[t]
    \centering
    \includegraphics[width=0.8\linewidth]{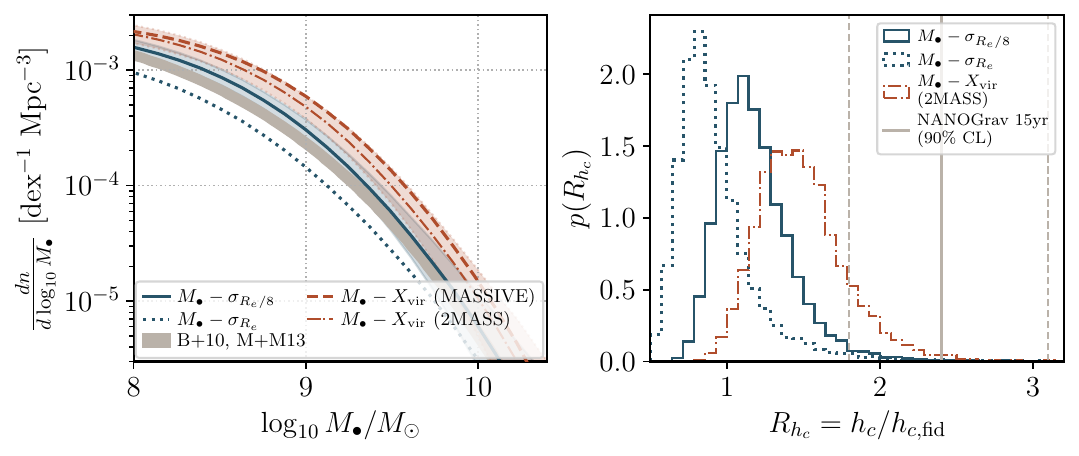}
    \caption{Black hole mass function on the left and characteristic strain values on the right. Blue and red lines correspond to inferences from this work based on $\sigma$ and $X_{\rm vir}$, while gray lines correspond to external measurements. On the left, the external measurements are the SDSS velocity dispersion function~\cite{bernardi_MF} (B+10) and the scaling relation from Ref.~\cite{mcconnell_ma} (M+M13). The blue solid line corresponds to the equivalent measurement using the 6dFGS catalog, while the dotted line includes the correction for the fact that BH samples measure $\sigma$ at the effective radius $R_e$, while MF samples (both 6dFGS and SDSS) measure at $R_e/8$. Dash-dotted and dashed red lines show the estimates based on $X_{\rm vir}$ using the 2MASS magnitude and radii and the the values corrected using the fit from the MASSIVE sample, respectively. On the right panel, the gray vertical lines show the value measured in the NANOGrav 15yr dataset~\cite{NANOGrav_stoc}.}
    \label{fig:BHMF_hc}
\end{figure*}

We measure the velocity dispersion function (VDF) following the methodology outlined in Sec.~\ref{subsec:MF}.  Our methodology first determines the $\sigma - X_{\rm vir}$ relation in the Fundamental Plane (FP) sample (6dFGSv) and then combines it with the measurement of the $X_{\rm vir}$ function in the full 6dFGS catalog. Mathematically, this can be expressed as
\begin{equation}
    \frac{dn}{d\log \sigma} = \int dX_{\rm vir}\ p(\log \sigma| X_{\rm vir})\frac{dn}{dX_{\rm vir}}.
    \label{eq:VDF_Xvir}
\end{equation}
The remarkable agreement between our VDF measurement and that of Ref.~\cite{bernardi_MF} using SDSS data is illustrated explicitly in Fig.~\ref{fig:VDF}. The agreement suggests a robust measurement of the VDF from galaxy surveys. 

Utilizing the FP relation from the FP sample renders us immune to potential errors arising from inconsistencies in the definition of $X_{\rm vir}$ between the black hole (BH) and mass function (MF) samples. Although the FP sample predominantly contains early-type galaxies, raising concerns about the universal applicability of the derived $X_{\rm vir}-\sigma$ relation, the strong agreement with Ref.~\cite{bernardi_MF} indicates any such effect is likely minor.

If the $p(\log \sigma| X_{\rm vir})$ inferred from the BH sample is used in Eq.~\ref{eq:VDF_Xvir} instead of that from the 6dFGSv, the dashed orange curve is obtained which is significantly higher than the estimate of Ref.~\cite{bernardi_MF}. Therefore, when assigning a velocity dispersion to each galaxy based on $X_{\rm vir}$, it appears more consistent to use the scaling relation derived from the FP sample. If we assumed instead that the BH sample FP relation was universally applicable and that there were no inconsistencies between BH and MF samples we would have to find an explanation for the disagreement.  We would have to either conclude that there is a mismatch in the definition of $\sigma$ between Ref.~\cite{bernardi_MF} and the FP sample (for example due to aperture corrections) or accept that Ref.~\cite{bernardi_MF}'s count is underestimated by roughly a factor of two.

\subsection{Mass Functions}\label{sec:MF}
To estimate the BH mass function, the straightforward approach is to count galaxies as a function of $X_{\rm vir}$ and use scaling relations to assign a BH to each galaxy in the catalog, just as we did for the VDF. We have seen, however, that there are inconsistencies between the BH and MF samples, shown by the fact that the velocity dispersion associated with a given value of $X_{\rm vir}$ is not the same in the two samples and that the fundamental planes are not equally aligned.

One source of systematics are differences in the value of $X_{\rm vir}$ that a given galaxy would have if part of the BH or MF samples due to observational differences. If these were the dominant source of systematics, one could get around it by using the velocity dispersion. The FP sample allows us to infer the velocity dispersion in the MF sample and the BH sample has both the velocity dispersion and black hole mass. Thus one is calibrating out potential differences in the meaning of $X_{\rm vir}$ in both samples, {\it ie.} calibrating $p(x|X)$. We will adopt this route by estimating the velocity dispersion in each galaxy as in the previous section and use the  $M_\bullet-\sigma$ relation from the BH sample to assign a BH mass based on this velocity dispersion. 

It could also be that the differences we saw in the velocity dispersion between BH and MF samples are the result of differences in selection. In that limit, $X_{\rm vir}$ has the same meaning in both samples and one can use the $M_\bullet-X_{\rm vir}$ relation in the BH sample to estimate the BH masses in the MF sample. We will also adopt this approach. In this scenario, a galaxy of a given $X_{\rm vir}$ in the MF sample would have a lower velocity dispersion than would be implied by the $\sigma-X_{\rm vir}$ relation of the BH sample but it would still have a BH of the same mass. However, a selection effect that changes the expected velocity dispersion for a given $X_{\rm vir}$ could also result in an additional shift in the black hole mass (for example, if the missing portion of the sample with lower $\sigma$ also has a correspondingly lower BH mass). This cannot be readily estimated and, ultimately, a complete local survey of BHs or at least one with very well-understood selection effects is the only way to make progress. Perhaps in the meantime one can take the difference in the mass functions we will find using these two routes as an estimate of potential systematic errors. 

Following the methodology outlined in Sec.~\ref{subsec:MF} we also measure the black hole mass function using $\sigma$ or $X_{\rm vir}$ as proxies for the black hole mass. When using $M_\bullet-\sigma$, we use the estimate of $\sigma$ discussed in the previous section. We also incorporate corrections to the observed magnitude, radius, and velocity dispersion. 

The gray band in the left panel of Fig.~\ref{fig:BHMF_hc} represents the mass function obtained by combining the velocity dispersion function (VDF) from SDSS measured in Ref.~\cite{bernardi_MF} with the $M_{\bullet}-\sigma$ relation from Ref.~\cite{mcconnell_ma}. Our equivalent measurement, using an independently derived VDF, is the solid blue line. While the local catalog we adopt from Ref.~\cite{R_van_den_Bosch2016} closely matches Ref.~\cite{mcconnell_ma}, our VDF measurement is derived through Eq.~\ref{eq:VDF_Xvir}, combining our measured $\sigma - X_{\rm vir}$ relation from the FP sample with the $X_{\rm vir}$ distribution in the full 6dFGS catalog.

\begin{figure*}[t]
    \centering
    \includegraphics[width=0.85\linewidth]{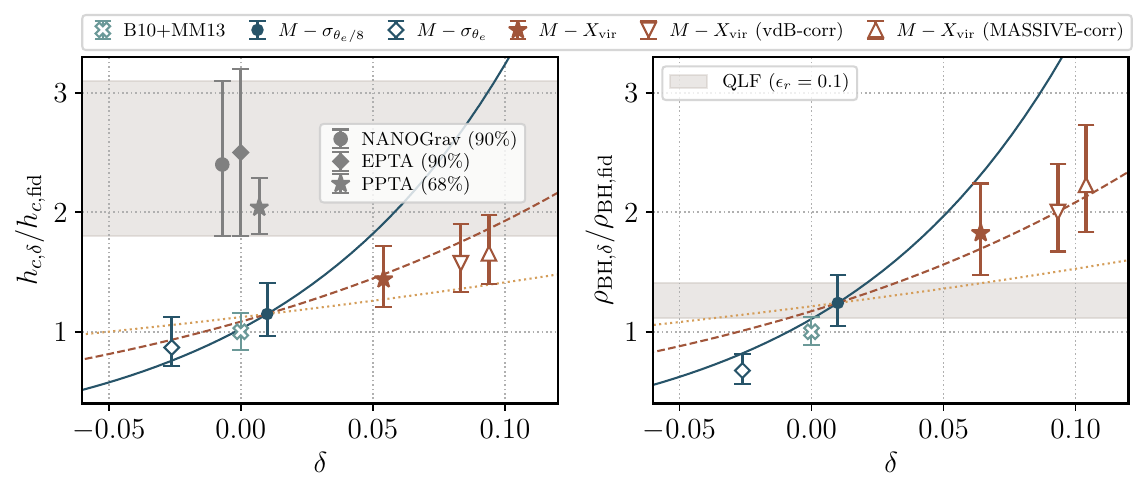}
    \caption{Characteristic strain normalized by the fiducial value (combining mass function and scaling relation from Refs.~\cite{bernardi_MF} and \cite{mcconnell_ma}) on the left, and the black hole mass density on the right. The light blue cross shows the fiducial value, the dark blue points are the measurements based on $\sigma$, and the red points are based on $X_{\rm vir}$. Different symbols correspond to different systematic corrections. The gray points and bands correspond to external measurements.}
    \label{fig:hc_summary}
\end{figure*}

There is evidence for systematic offsets in the mass proxies between the BH and FP/MF samples. In the velocity dispersion case, BH catalogs typically report $\sigma$ at the effective radius $R_e$, while the mass function catalogs report closer to the center, at $R_e/8$. Since velocity dispersion profiles for the most massive galaxies are consistently found to rise at the center, this correction strictly reduces the inferred BH mass function. The case shown in Figs.~\ref{fig:BHMF_hc} and \ref{fig:hc_summary} corresponds to the commonly adopted value for the slope of the $\sigma$ profile of $\gamma_{\rm ap} = -0.04$. The fiducial mass function in gray from combining Refs.~\cite{bernardi_MF} and \cite{mcconnell_ma} suffers from the same offset, since SDSS also reports $\sigma$ at $R_e/8$, and would be scaled by the same factor.

While magnitude and radius corrections can also have a significant impact, they are not calibrated in the relevant parameter range. Both of the calibrated corrections for the offsets in magnitude and radius are calibrated in the local sample only, resulting in a linear relation that is extrapolated beyond its measured range of values when applied to the mass function sample. We nevertheless include them in the results, but emphasize its limitations.

We summarize the resulting characteristic strain values from all estimates of the SMBH mass function described in this work in Fig.~\ref{fig:hc_summary} as the colored points, along with the measurements reported by PTA collaborations (\cite{NANOGrav_stoc, EPTA, PPTA}) in gray. The location of the PTA data points along the x-axis is arbitrary, while the values of $\delta$ for the colored points were obtained as follows. The fiducial value combining Refs.\cite{mcconnell_ma} and \cite{bernardi_MF} is defined as the zero point and $\delta$ for the equivalent estimate derived in this work (filled blue dot) is assumed to be the average shift measured by Ref.~\cite{6dFGSv_FP} between the 6dFGSv sample and SDSS. The other values of $\delta$ are computed relative to this: when including aperture correction in $\sigma$ we adopt $\delta=-0.04\log 8$, and the red points correspond to the difference between $X_{\rm vir}$ in the BH and FP samples computed using the corrections reported in each paper at the peak velocity dispersion $\sigma_{\rm peak}$. That is, the value of $\delta$ for the red points follows directly from the gap between the BH and FP scaling relations shown in Fig.~\ref{fig:sig_LR_L} and is given by the difference in $X_{\rm vir}$ at $\sigma_{\rm peak}$. The lines show the characteristic strain as a function of the shift in galaxy property (from Eq.~\ref{eq:h2c_shift}), assuming $b_{\bullet}=5, 2.5,$ and $1$, and are normalized to the blue point. The measured values for the scaling relation coefficients for $M_{\bullet}-\sigma$ and $M_{\bullet}-X_{\rm vir}$ are shown in Figs.~\ref{fig:M-sigma} and \ref{fig:M-Xvir} of App.~\ref{app:M-sigma}.

The general agreement between the colored points and the lines shows that the variations in GWB amplitude are consistent with the observed offsets in the galaxy properties. For instance, if the velocity dispersion functions were inconsistent between SDSS and 6dFGS, or the scaling relations were severely mismodeled, the points would significantly deviate from the curves. Differences in these predictions are therefore not related directly to black hole masses themselves. Beyond this discrepancy (either due to observation systematics or selection effects), the only other way to produce inconsistent mass function measurements from the same data set (i.e., between red and blue points) would be to mismodel the scaling relation. We explore this in App.~\ref{app:M-sigma}, focusing on potential deviations of the $M_{\bullet}-\sigma$ relation from a simple power law, but show that there current data suggests that this is not a significant effect.

Since the VDF inferred from different surveys are in such good agreement, it is reasonable to suppose that, if the mass function inferred from $\sigma$ is biased low, then the origin is likely due to inconsistencies between the BH and MF catalogs, and not due to the galaxy count. Furthermore, given that the dominant observational systematic error are aperture corrections and evidence suggests that this only reduces the MF estimate (see Sec.~\ref{sec:vdisp_corr}), the origin of the underestimation would most likely be from selection effects. To estimate the magnitude from selection effects, we follow an identical discussion to Sec.~\ref{subsec:sig_L_R} and App.~\ref{app:FP}. We may suppose that the BH mass depends on a second quantity $Y$, in addition to $\sigma$, and that the distribution of $Y$ is biased in the BH sample relative to the MF catalog. The inferred intrinsic scatter $\varepsilon_{\bullet}$ would therefore be partially absorbing the range in $Y$ and the shift from selection effects is some fraction of $\varepsilon_{\bullet}$. From Refs.~\cite{mcconnell_ma} and \cite{R_van_den_Bosch2016} (also shown in App.~\ref{app:M-sigma}), the scatter is typically $\varepsilon_{\bullet} = 0.4$. In order to bridge the gap between the dark blue point and the PTA measurements shown in Fig.~\ref{fig:hc_summary}, a shift larger than $1\sigma$ is required --- a shift $\delta=0.4$ from selection effects corresponds to an $h_c$ larger by a factor of 2.15.

The results for the black hole mass density are derived in an equivalent manner to $h_c$, but with $\gamma=1$ instead of $5/6$. The gray band is obtained by integrating the quasar luminosity function from Ref.~\cite{shen_2020} (see, e.g., Ref.~\cite{Sato-Polito:2025ivz} for further details). The range corresponds to 1$\sigma$ uncertainties in the quasar luminosity function (QLF) and the mass density is obtained by fixing the radiative efficiency to a representative value $\epsilon_r=0.1$. Of course, it is possible for the radiative efficiency to have a different value, but we show $\epsilon_r=0.1$ as a guide.

\section{Conclusion}
The SMBH mass function is the most important quantity for the theoretical interpretation of the GW signal reported by PTA collaborations. From it, the amplitude of the GW background can be predicted, including a strict upper limit that is independent of the merger history~\cite{Phinney:2001di, Sato-Polito:2023gym}. We explored potential sources of uncertainty in current estimates of the mass function, using the 6dFGS and 2MASS catalogs as testbeds. In particular, we focus on $\sigma$, $L$, and the virial combination $X_{\rm vir}$ as proxies for the black hole mass, and investigate the impact that internal inconsistencies between galaxy properties can produce on the GW signal. 

We find that shifts in the mass proxy that are relatively small when compared to systematic corrections found empirically in the literature can lead to changes in the SMBH mass function comparable to or larger than the uncertainties currently reported by PTAs, highlighting the importance of the GW-based inference. This also suggests that neglecting to homogenize the mass proxy definition between the BH and MF catalogs entirely may lead to very large errors on the predicted mass function. 

The tolerance for error $\delta$ in the mass proxy is a strong function of the slope of the $M_{\bullet} - X$ relation ($h_c\propto 10^{\frac{5}{6} b_{\bullet} \delta}$). While velocity dispersions appear to be more reliably measured than $L$ and $R$, they also have the steepest slope. However, the only known corrections strictly decrease the mass function estimate. On the other hand, $L$ and $R$ are subject to much larger systematic corrections which are, themselves, not particularly well measured in the regimes relevant to this study. In spite of this, $\sigma$ and $X_{\rm vir}$-based estimates appear to be relatively consistent.

While $X_{\rm vir}$ and $L$ initially appear to produce different estimates of the SMBH mass function, we show that this difference can be understood as a difference in surface brightness of the samples. The relations between $\sigma-X_{\rm vir}$ and $\sigma-L$ correspond to different projections of the fundamental plane and the scatter of the $\sigma-L$ relation absorbs the range of surface brightness values. While we do not discuss stellar masses, it is plausible that similar effects are present in mass function estimates based on it --- both the systematic shifts in $M_*$ across BH and MF catalogs, and potentially a projection analogous to the $\sigma-L$ case. This discussion also provides a general account for the potential impact of selection effects on any BH mass function measurement based on scaling relations.


Although the $\sigma$ and $X_{\rm vir}$ based estimates of the background are relatively consistent, a small discrepancy still remains. While we show that the origin of this discrepancy is the difference in the fundamental plane in the local and mass function catalogs, cause of this mismatch is unclar. If the fundamental plane measured in the local sample were applied to the mass function sample, the VDF would be very different from the one measured directly so this this seem to imply that the local scaling relations cannot be applied unchanged to the MF sample. It is plausible that small shifts in magnitudes or radii when measured in each sample are the cause of the difference. But we also find some evidence that selection effects may be playing a role, which once corrected would lead to a lower $\sigma$ for a given $X_{\rm vir}$. In either case the VDF would be left more or less unchanged relative to the fiducial value. To obtain a larger predicted stochastic background, consistent with the PTA measurements, one would need to assume that the selection differences lead to a black hole mass larger by about a factor of two for a given $\sigma$, equivalent to a shift equal to $1\sigma$ of the intrinsic scatter. 

\section*{Acknowledgments}
We would like to thank Eliot Quataert, Jenny Greene, and Michael Strauss for helpful discussions. MZ is supported by NSF 2209991 and NSF-BSF 2207583. GSP is supported by the Friends of the Institute for Advanced Study Fund.

\appendix
\section{Fundamental plane and $\sigma-L$}\label{app:FP}

\begin{figure}[t]
    \centering
    \includegraphics[width=0.9\linewidth]{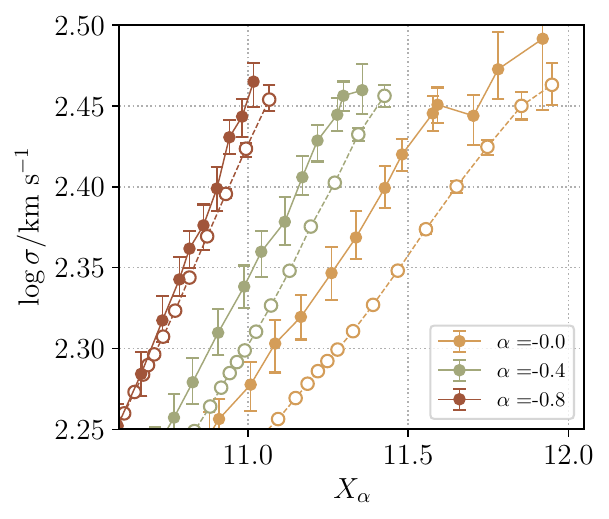}
    \caption{Relation between $\sigma$ and $X_{\alpha} = \log L + \alpha \log R$ for different values of $\alpha$. The filled and unfilled dots correspond to the BH and FP samples, respectively.}
    \label{fig:sig_X_alpha}
\end{figure}

As argued in the main text, the relations $\sigma-X_{\rm vir}$ and $\sigma-L$ are particular projections of the fundamental plane. We show in Fig.~\ref{fig:sig_X_alpha} the relation between $\sigma-X_{\alpha}$ with $X_{\alpha}\equiv \log L + \alpha \log R$, where $\alpha=0$ corresponds to the $\sigma-L$ relation and $\alpha=\alpha_{\rm vir}\sim -0.8$ is the edge-on view of the fundamental plane in the FP sample. The reduction in the gap between filled and unfilled dots shows that the discrepancy between the $\sigma$ assigned to a given $X_{\alpha}$ in the BH and FP catalogs are reduced as $\alpha$ approaches the fundamental plane value. The remaining gap seen in the red points is the same as shown in the left panel of Fig.~\ref{fig:sig_LR_L}, and is a result of the fundamental planes in the BH and FP samples being different, as discussed in Sec.~\ref{subsec:sig_L_R}.

To understand this effect more quantitatively, suppose that galaxies have a set of properties $X$, $Y$, and $Z$, that satisfy
\begin{equation}
    Z = a + bX + cY +n_{XY},
\end{equation}
where $n_{XY} \sim \mathcal{N}(\mu=0, \sigma=\varepsilon_{XY})$ is an intrinsic noise term with scatter $\varepsilon_{XY}$. If we then look only at the relation 
\begin{equation}
    Z = a' + b'X + n_{X},
    \label{eq:z_scaling}
\end{equation}
we will have absorbed the dependence on $Y$ into the parameters $a'$ and $n_X$. Suppose that the parameter $Y$ is itself distributed in the particular sample in question as a gaussian $Y\sim \mathcal{N}(\mu=\avg{Y}, \sigma=\sigma_{Y})$, then $a'=a+c\avg{Y}$ and $\varepsilon_X = \sqrt{\sigma^2_{Y} + \varepsilon^2_{XY}}$. If the values of $Y$ are different in the BH and FP/MF samples, one finds different scaling relations when fitting Eq.~\ref{eq:z_scaling}. We also note that, while in principle any scaling relation used to predict $Z$ should yield the same answer, a lower scatter leaves less room for potential inconsistencies between different samples, since here $\varepsilon_X$ is absorbing the range in $Y$ (given by $\sigma_{Y}$).

\begin{figure}[t]
    \centering
    \includegraphics[width=0.9\linewidth]{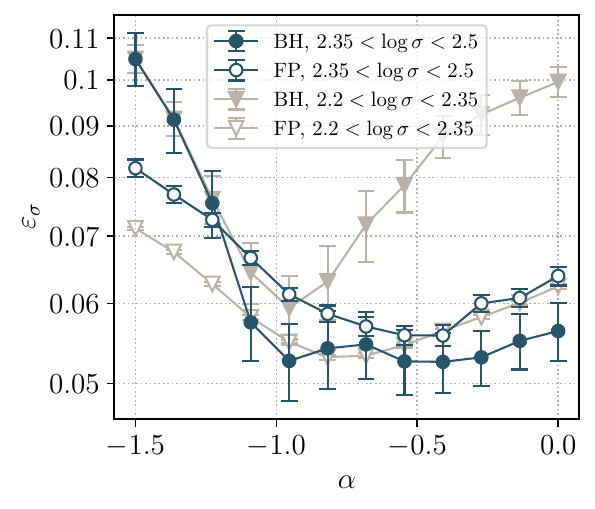}
    \caption{Inferred intrinsic scatter in $\sigma$ as a function of the virial parameter $\alpha$ in the BH (filled) and FP (unfilled) dots. We also divide between objects with low (light gray) and high (dark blue) velocity dispersions.}
    \label{fig:scatter_alpha}
\end{figure}

Fig.~\ref{fig:scatter_alpha} shows the inferred intrinsic scatter as a function of the value of $\alpha$ in the $X_{\alpha}$ definition. Filled and open symbols correspond to the results in the BH and FP samples, while dark blue and light gray are galaxies with high and low velocity dispersions, respectively. The dark blue therefore corresponds to the regime relevant to GW estimates. The smallest intrinsic scatter is found for $\alpha$ between -0.5 and -1 in all cases, reflecting the edge-on view of the fundamental plane. The large increase in scatter in the BH sample in the low $\sigma$ range is a result of population differences between spiral and elliptical galaxies which also get absorbed in the scatter. Once the virial combination is computed or one focuses on the high $\sigma$ regime, the values for the scatter becomes consistent across catalogs and $\sigma$ ranges.

\section{Curvature in $M_{\bullet}-\sigma$}
\begin{figure*}
    \centering
    \includegraphics[width=0.7\linewidth]{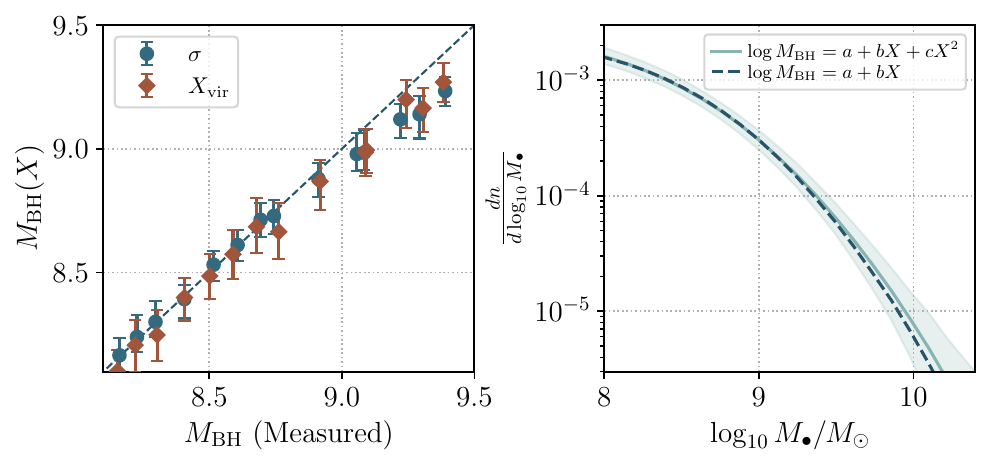}
    \caption{Impact of curvature in the $M_{\bullet}-\sigma$ relation on the SMBH mass function. The left panel shows the relation between the measured SMBH mass, and the mean mass assigned from the scaling relation fit by a linear relation. The blue points correspond to the value assigned based on $\sigma$ and the red are based on $X_{\rm vir}$, while the dashed line shows $y=x$. The panel on the right is the SMBH mass function inferred when fitting a linear (dashed dark blue) or quadratic (solid light blue) relation to $M_{\bullet}-\sigma$ for the $\sigma$-based estimate of the mass function. The dashed dark blue line is therefore identical to the solid blue line of Fig.~\ref{fig:BHMF_hc}.}
    \label{fig:mbh_quadratic}
\end{figure*}

In Sec.~\ref{sec:intro}, we argued that mismodeling of scaling relations could bias the inference of the SMBH mass function, but, if such an effect were significant, it would be fairly self-evident. One regime in which the scaling relations might be mismodeled is in the high-velocity dispersion end of $M_{\bullet}-\sigma$. Although the sample size in the massive regime is still relatively low, the $M-\sigma$ does somewhat appear to steepen for $\sigma\gtrsim 275$km s$^{-1}$\cite{kormendy_ho, mcconnell_ma}, as a result of the galaxy population ``saturating" at high $\sigma$ \cite{2013ApJ...769L...5K, Thomas:2013bga, 2012MNRAS.424..224H}. We show the impact of this effect on the predicted black hole mass in Fig.~\ref{fig:mbh_quadratic}. Similarly to Fig.~\ref{fig:sig_LR_L}, the points with error bars in the figure are obtained from bootstraping, where in each iteration a linear scaling relation is fit between $\log M_{\bullet}$, and $\log \sigma$ or $X_{\rm vir}$. Each galaxy in the local sample is assigned a mean black hole mass based on the scaling relation (shown in the x-axis), and we show the relation between the mass from the fit and the true measured value in bins, which have a width 5 times the bin separation. The slight underprediction of the SMBH mass (y-axis) relative to the measured value (x-axis) is the effect described above.

From previous works in the literature~\cite{2006MNRAS.365.1082W, 2009ApJ...698..198G, mcconnell_ma}, we expect the effect on the GW amplitude to be small. For example, using the fiducial SDSS VDF \cite{bernardi_MF} and comparing the resulting $h_c$ for the best-fit linear or quadratic scaling relations in Ref.~\cite{mcconnell_ma}, one finds that $h_c$ shifts by only 5\%. We also show the result explicitly in the 6dFGS/2MASS data set, and include a quadratic relation between $\log \sigma$ and $\log M_{\bullet}$. We follow an identical approach to that outlined in Sec.~\ref{subsec:MF} and show the resulting mass functions in Fig.~\ref{fig:mbh_quadratic}. The small increase in BH abundance at high masses is a consequence of the ``overmassive'' black holes that exceed the linear $M_{\bullet}-\sigma$ relation at high $\sigma$. The median of the bootstrap samples indicates an increase in $h_c$ of $\sim 15\%$. This shift is about the same as the difference between the fiducial estimate (light blue cross in Fig.~\ref{fig:hc_summary}) and the $\sigma$-based result obtained in this work (dark blue dot), and we therefore consider it a sub-dominant effect. 

\section{BH scaling relations}\label{app:M-sigma}
\begin{figure*}
    \centering
    \includegraphics[width=0.9\linewidth]{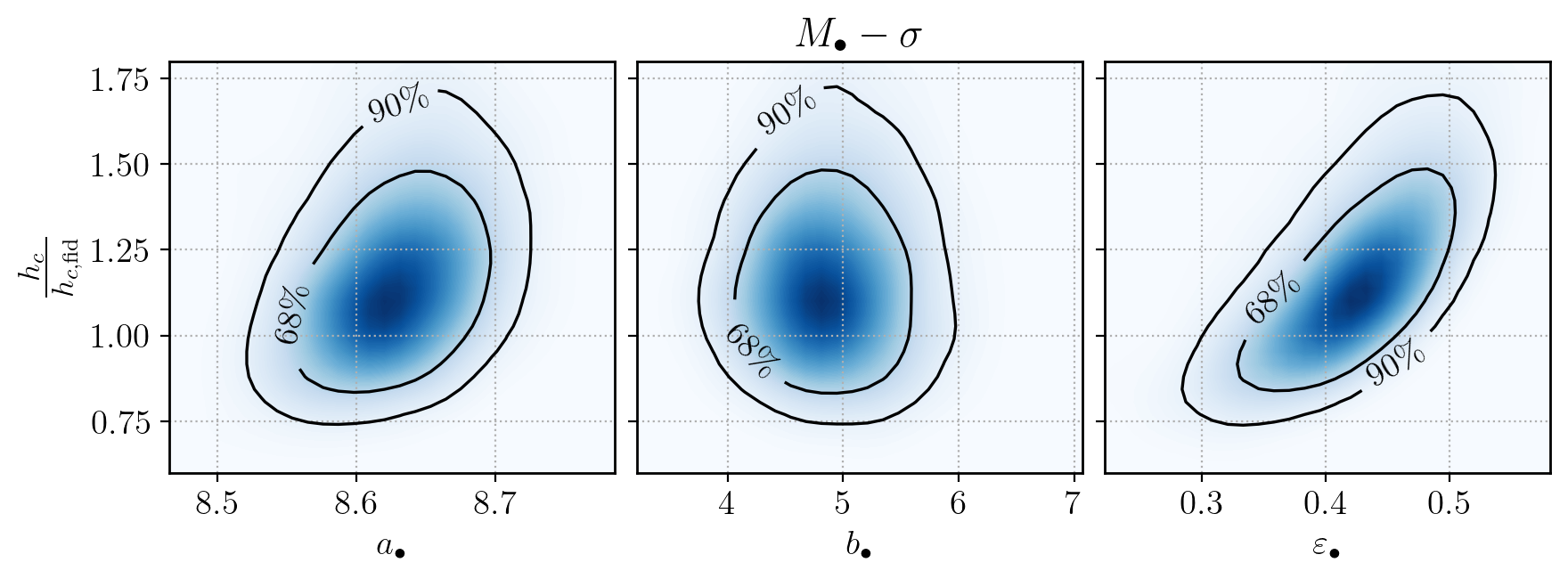}
    \caption{Histogram of scaling relation coefficients for $M_{\bullet}-\sigma$ and $h_c$ in bootstrap samples.}
    \label{fig:M-sigma}
\end{figure*}

\begin{figure*}
    \centering
    \includegraphics[width=0.9\linewidth]{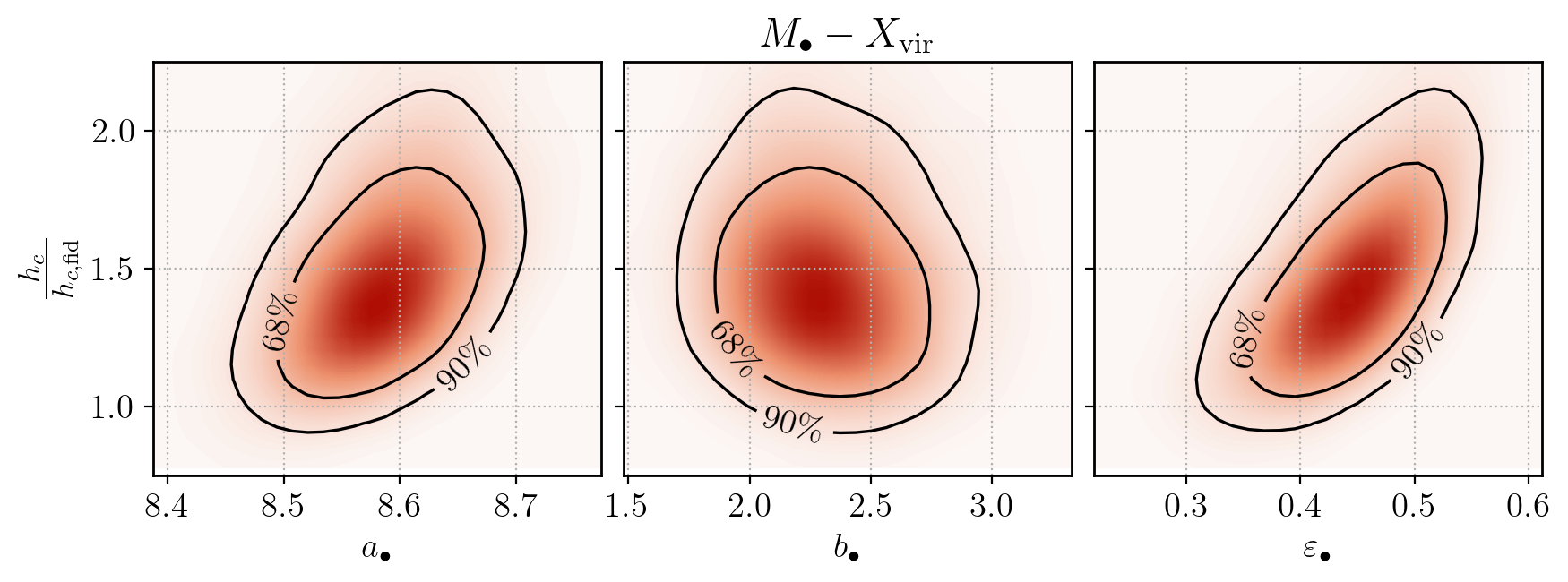}
    \caption{Equivalent to Fig.~\ref{fig:M-sigma}, but for $M_{\bullet}-X_{\rm vir}$ using the values of L and R reported by 2MASS.}
    \label{fig:M-Xvir}
\end{figure*}

For the sake of completeness, we show in this appendix the black hole scaling relations for $\sigma$ and $X_{\rm vir}$. Figs~\ref{fig:M-sigma} and \ref{fig:M-Xvir} are 2D histograms of the $10^4$ bootstrap iterations, with the scaling relation coefficients (defined in Eq.~\ref{eq:M-X}) on the x-axis and the corresponding $h_c$, normalized by the fiducial result (combining the VDF of Ref.~\cite{bernardi_MF} with the $M_{\bullet}-\sigma$ from \cite{mcconnell_ma}), on the y-axis. The pivot for the linear fit is chosen such that $a_{\bullet}$ and $b_{\bullet}$ are uncorrelated, which are $\log\sigma_{\rm pivot}=2.39$ and $X_{\rm pivot}=10.76$. Fig.~\ref{fig:M-Xvir} uses the uncorrected 2MASS L and R values.

\bibliography{ref.bib}

\providecommand{\href}[2]{#2}\begingroup\raggedright\begin{thebibliography}{10}

\bibitem{EPTA}
{\bfseries EPTA} Collaboration, J.~Antoniadis {\em et~al.}, ``{The second data release from the European Pulsar Timing Array III. Search for gravitational wave signals},'' \href{http://arxiv.org/abs/2306.16214}{{\ttfamily arXiv:2306.16214 [astro-ph.HE]}}.

\bibitem{NANOGrav_stoc}
{\bfseries NANOGrav} Collaboration, G.~Agazie {\em et~al.}, ``{The NANOGrav 15 yr Data Set: Evidence for a Gravitational-wave Background},'' \href{http://dx.doi.org/10.3847/2041-8213/acdac6}{{\em Astrophys. J. Lett.} {\bfseries 951} no.~1, (2023) L8}, \href{http://arxiv.org/abs/2306.16213}{{\ttfamily arXiv:2306.16213 [astro-ph.HE]}}.

\bibitem{PPTA}
D.~J. Reardon {\em et~al.}, ``{Search for an Isotropic Gravitational-wave Background with the Parkes Pulsar Timing Array},'' \href{http://dx.doi.org/10.3847/2041-8213/acdd02}{{\em Astrophys. J. Lett.} {\bfseries 951} no.~1, (2023) L6}, \href{http://arxiv.org/abs/2306.16215}{{\ttfamily arXiv:2306.16215 [astro-ph.HE]}}.

\bibitem{CPTA}
H.~Xu {\em et~al.}, ``{Searching for the Nano-Hertz Stochastic Gravitational Wave Background with the Chinese Pulsar Timing Array Data Release I},'' \href{http://dx.doi.org/10.1088/1674-4527/acdfa5}{{\em Res. Astron. Astrophys.} {\bfseries 23} no.~7, (2023) 075024}, \href{http://arxiv.org/abs/2306.16216}{{\ttfamily arXiv:2306.16216 [astro-ph.HE]}}.

\bibitem{IPTA}
{\bfseries International Pulsar Timing Array} Collaboration, G.~Agazie {\em et~al.}, ``{Comparing recent PTA results on the nanohertz stochastic gravitational wave background},'' \href{http://arxiv.org/abs/2309.00693}{{\ttfamily arXiv:2309.00693 [astro-ph.HE]}}.

\bibitem{MPTA}
M.~T. {Miles}, R.~M. {Shannon}, D.~J. {Reardon}, M.~{Bailes}, D.~J. {Champion}, M.~{Geyer}, {\em et~al.}, ``{The MeerKAT Pulsar Timing Array: the 4.5-yr data release and the noise and stochastic signals of the millisecond pulsar population},'' \href{http://dx.doi.org/10.1093/mnras/stae2572}{{\em \mnras} {\bfseries 536} no.~2, (Jan., 2025) 1467--1488}, \href{http://arxiv.org/abs/2412.01148}{{\ttfamily arXiv:2412.01148 [astro-ph.HE]}}.

\bibitem{Burke-Spolaor:2018bvk}
S.~Burke-Spolaor {\em et~al.}, ``{The Astrophysics of Nanohertz Gravitational Waves},'' \href{http://dx.doi.org/10.1007/s00159-019-0115-7}{{\em Astron. Astrophys. Rev.} {\bfseries 27} no.~1, (2019) 5}, \href{http://arxiv.org/abs/1811.08826}{{\ttfamily arXiv:1811.08826 [astro-ph.HE]}}.

\bibitem{1995ARA&A..33..581K}
J.~{Kormendy} and D.~{Richstone}, ``{Inward Bound---The Search For Supermassive Black Holes In Galactic Nuclei},'' \href{http://dx.doi.org/10.1146/annurev.aa.33.090195.003053}{{\em \araa} {\bfseries 33} (Jan., 1995) 581}.

\bibitem{1998Natur.395A..14R}
D.~{Richstone}, E.~A. {Ajhar}, R.~{Bender}, G.~{Bower}, A.~{Dressler}, S.~M. {Faber}, {\em et~al.}, ``{Supermassive black holes and the evolution of galaxies.},'' \href{http://dx.doi.org/10.48550/arXiv.astro-ph/9810378}{{\em \nat} {\bfseries 385} no.~6701, (Oct., 1998) A14}, \href{http://arxiv.org/abs/astro-ph/9810378}{{\ttfamily arXiv:astro-ph/9810378 [astro-ph]}}.

\bibitem{1980Natur.287..307B}
M.~C. {Begelman}, R.~D. {Blandford}, and M.~J. {Rees}, ``{Massive black hole binaries in active galactic nuclei},'' \href{http://dx.doi.org/10.1038/287307a0}{{\em \nat} {\bfseries 287} no.~5780, (Sept., 1980) 307--309}.

\bibitem{kormendy_ho}
J.~{Kormendy} and L.~C. {Ho}, ``{Coevolution (Or Not) of Supermassive Black Holes and Host Galaxies},'' \href{http://dx.doi.org/10.1146/annurev-astro-082708-101811}{{\em \araa} {\bfseries 51} no.~1, (Aug., 2013) 511--653}, \href{http://arxiv.org/abs/1304.7762}{{\ttfamily arXiv:1304.7762 [astro-ph.CO]}}.

\bibitem{mcconnell_ma}
N.~J. {McConnell} and C.-P. {Ma}, ``{Revisiting the Scaling Relations of Black Hole Masses and Host Galaxy Properties},'' \href{http://dx.doi.org/10.1088/0004-637X/764/2/184}{{\em \apj} {\bfseries 764} no.~2, (Feb., 2013) 184}, \href{http://arxiv.org/abs/1211.2816}{{\ttfamily arXiv:1211.2816 [astro-ph.CO]}}.

\bibitem{2004MNRAS.351..169M}
A.~{Marconi}, G.~{Risaliti}, R.~{Gilli}, L.~K. {Hunt}, R.~{Maiolino}, and M.~{Salvati}, ``{Local supermassive black holes, relics of active galactic nuclei and the X-ray background},'' \href{http://dx.doi.org/10.1111/j.1365-2966.2004.07765.x}{{\em \mnras} {\bfseries 351} no.~1, (June, 2004) 169--185}, \href{http://arxiv.org/abs/astro-ph/0311619}{{\ttfamily arXiv:astro-ph/0311619 [astro-ph]}}.

\bibitem{vika2009}
M.~{Vika}, S.~P. {Driver}, A.~W. {Graham}, and J.~{Liske}, ``{The Millennium Galaxy Catalogue: the M$_{bh}$-L$_{spheroid}$ derived supermassive black hole mass function},'' \href{http://dx.doi.org/10.1111/j.1365-2966.2009.15544.x}{{\em \mnras} {\bfseries 400} no.~3, (Dec., 2009) 1451--1460}, \href{http://arxiv.org/abs/0908.2102}{{\ttfamily arXiv:0908.2102 [astro-ph.CO]}}.

\bibitem{shankar2009}
F.~{Shankar}, D.~H. {Weinberg}, and J.~{Miralda-Escud{\'e}}, ``{Self-Consistent Models of the AGN and Black Hole Populations: Duty Cycles, Accretion Rates, and the Mean Radiative Efficiency},'' \href{http://dx.doi.org/10.1088/0004-637X/690/1/20}{{\em \apj} {\bfseries 690} no.~1, (Jan., 2009) 20--41}, \href{http://arxiv.org/abs/0710.4488}{{\ttfamily arXiv:0710.4488 [astro-ph]}}.

\bibitem{shankar2013}
F.~{Shankar}, ``{Black hole demography: from scaling relations to models},'' \href{http://dx.doi.org/10.1088/0264-9381/30/24/244001}{{\em Classical and Quantum Gravity} {\bfseries 30} no.~24, (Dec., 2013) 244001}, \href{http://arxiv.org/abs/1307.3289}{{\ttfamily arXiv:1307.3289 [astro-ph.CO]}}.

\bibitem{Phinney:2001di}
E.~S. Phinney, ``{A Practical theorem on gravitational wave backgrounds},'' \href{http://arxiv.org/abs/astro-ph/0108028}{{\ttfamily arXiv:astro-ph/0108028}}.

\bibitem{Sato-Polito:2023gym}
G.~Sato-Polito, M.~Zaldarriaga, and E.~Quataert, ``{Where are the supermassive black holes measured by PTAs?},'' \href{http://dx.doi.org/10.1103/PhysRevD.110.063020}{{\em Phys. Rev. D} {\bfseries 110} no.~6, (2024) 063020}, \href{http://arxiv.org/abs/2312.06756}{{\ttfamily arXiv:2312.06756 [astro-ph.CO]}}.

\bibitem{Sato-Polito:2024lew}
G.~Sato-Polito and M.~Zaldarriaga, ``{Distribution of the gravitational-wave background from supermassive black holes},'' \href{http://dx.doi.org/10.1103/PhysRevD.111.023043}{{\em Phys. Rev. D} {\bfseries 111} no.~2, (2025) 023043}, \href{http://arxiv.org/abs/2406.17010}{{\ttfamily arXiv:2406.17010 [astro-ph.CO]}}.

\bibitem{Sato-Polito:2025ivz}
G.~Sato-Polito, M.~Zaldarriaga, and E.~Quataert, ``{Evolution of SMBHs in light of PTA measurements: implications for growth by mergers and accretion},'' \href{http://arxiv.org/abs/2501.09786}{{\ttfamily arXiv:2501.09786 [astro-ph.CO]}}.

\bibitem{Lamb_free_spec}
W.~G. Lamb, S.~R. Taylor, and R.~van Haasteren, ``{Rapid refitting techniques for Bayesian spectral characterization of the gravitational wave background using pulsar timing arrays},'' \href{http://dx.doi.org/10.1103/PhysRevD.108.103019}{{\em Phys. Rev. D} {\bfseries 108} no.~10, (2023) 103019}, \href{http://arxiv.org/abs/2303.15442}{{\ttfamily arXiv:2303.15442 [astro-ph.HE]}}.

\bibitem{2007ApJ...662..808L}
T.~R. {Lauer}, S.~M. {Faber}, D.~{Richstone}, K.~{Gebhardt}, S.~{Tremaine}, M.~{Postman}, {\em et~al.}, ``{The Masses of Nuclear Black Holes in Luminous Elliptical Galaxies and Implications for the Space Density of the Most Massive Black Holes},'' \href{http://dx.doi.org/10.1086/518223}{{\em \apj} {\bfseries 662} no.~2, (June, 2007) 808--834}, \href{http://arxiv.org/abs/astro-ph/0606739}{{\ttfamily arXiv:astro-ph/0606739 [astro-ph]}}.

\bibitem{tundo2007}
E.~{Tundo}, M.~{Bernardi}, J.~B. {Hyde}, R.~K. {Sheth}, and A.~{Pizzella}, ``{On the Inconsistency between the Black Hole Mass Function Inferred from M$_{{\textbullet}}$-{\ensuremath{\sigma}} and M$_{{\textbullet}}$-L Correlations},'' \href{http://dx.doi.org/10.1086/518225}{{\em \apj} {\bfseries 663} no.~1, (July, 2007) 53--60}, \href{http://arxiv.org/abs/astro-ph/0609297}{{\ttfamily arXiv:astro-ph/0609297 [astro-ph]}}.

\bibitem{2007ApJ...660..267B}
M.~{Bernardi}, R.~K. {Sheth}, E.~{Tundo}, and J.~B. {Hyde}, ``{Selection Bias in the M$_{{\textbullet}}$-{\ensuremath{\sigma}} and M$_{{\textbullet}}$-L Correlations and Its Consequences},'' \href{http://dx.doi.org/10.1086/512719}{{\em \apj} {\bfseries 660} no.~1, (May, 2007) 267--275}, \href{http://arxiv.org/abs/astro-ph/0609300}{{\ttfamily arXiv:astro-ph/0609300 [astro-ph]}}.

\bibitem{2016MNRAS.460.3119S}
F.~{Shankar}, M.~{Bernardi}, R.~K. {Sheth}, L.~{Ferrarese}, A.~W. {Graham}, G.~{Savorgnan}, {\em et~al.}, ``{Selection bias in dynamically measured supermassive black hole samples: its consequences and the quest for the most fundamental relation},'' \href{http://dx.doi.org/10.1093/mnras/stw678}{{\em \mnras} {\bfseries 460} no.~3, (Aug., 2016) 3119--3142}, \href{http://arxiv.org/abs/1603.01276}{{\ttfamily arXiv:1603.01276 [astro-ph.GA]}}.

\bibitem{Shankar:2025unm}
F.~Shankar {\em et~al.}, ``{Probing the co-evolution of SMBHs and their hosts from scaling relations pairwise residuals: dominance of stellar velocity dispersion and host halo mass},'' \href{http://arxiv.org/abs/2505.02920}{{\ttfamily arXiv:2505.02920 [astro-ph.GA]}}.

\bibitem{Bernardi_LMfunc_2013}
M.~Bernardi, A.~Meert, R.~K. Sheth, V.~Vikram, M.~Huertas-Company, S.~Mei, and F.~Shankar, ``{The massive end of the luminosity and stellar mass functions: Dependence on the fit to the light profile},'' \href{http://dx.doi.org/10.1093/mnras/stt1607}{{\em Mon. Not. Roy. Astron. Soc.} {\bfseries 436} (2013) 697}, \href{http://arxiv.org/abs/1304.7778}{{\ttfamily arXiv:1304.7778 [astro-ph.CO]}}.

\bibitem{DSouza_2015}
R.~{D'Souza}, S.~{Vegetti}, and G.~{Kauffmann}, ``{The massive end of the stellar mass function},'' \href{http://dx.doi.org/10.1093/mnras/stv2234}{{\em \mnras} {\bfseries 454} no.~4, (Dec., 2015) 4027--4036}, \href{http://arxiv.org/abs/1509.07418}{{\ttfamily arXiv:1509.07418 [astro-ph.GA]}}.

\bibitem{Leja_2020}
J.~{Leja}, J.~S. {Speagle}, B.~D. {Johnson}, C.~{Conroy}, P.~{van Dokkum}, and M.~{Franx}, ``{A New Census of the 0.2 < z < 3.0 Universe. I. The Stellar Mass Function},'' \href{http://dx.doi.org/10.3847/1538-4357/ab7e27}{{\em \apj} {\bfseries 893} no.~2, (Apr., 2020) 111}, \href{http://arxiv.org/abs/1910.04168}{{\ttfamily arXiv:1910.04168 [astro-ph.GA]}}.

\bibitem{Liepold_2024}
E.~R. Liepold and C.-P. Ma, ``{Big Galaxies and Big Black Holes: The Massive Ends of the Local Stellar and Black Hole Mass Functions and the Implications for Nanohertz Gravitational Waves},'' \href{http://dx.doi.org/10.3847/2041-8213/ad66b8}{{\em Astrophys. J. Lett.} {\bfseries 971} no.~2, (2024) L29}, \href{http://arxiv.org/abs/2407.14595}{{\ttfamily arXiv:2407.14595 [astro-ph.GA]}}.

\bibitem{bernardi_MF}
M.~{Bernardi}, F.~{Shankar}, J.~B. {Hyde}, S.~{Mei}, F.~{Marulli}, and R.~K. {Sheth}, ``{Galaxy luminosities, stellar masses, sizes, velocity dispersions as a function of morphological type},'' \href{http://dx.doi.org/10.1111/j.1365-2966.2010.16425.x}{{\em \mnras} {\bfseries 404} no.~4, (June, 2010) 2087--2122}, \href{http://arxiv.org/abs/0910.1093}{{\ttfamily arXiv:0910.1093 [astro-ph.CO]}}.

\bibitem{2010ApJ...711L.108B}
D.~{Batcheldor}, ``{The M $_{{\textbullet}}$-{\ensuremath{\sigma}}$_{*}$ Relation Derived from Sphere of Influence Arguments},'' \href{http://dx.doi.org/10.1088/2041-8205/711/2/L108}{{\em \apjl} {\bfseries 711} no.~2, (Mar., 2010) L108--L111}, \href{http://arxiv.org/abs/1002.1705}{{\ttfamily arXiv:1002.1705 [astro-ph.CO]}}.

\bibitem{2023MNRAS.518.1352S}
N.~{Sahu}, A.~W. {Graham}, and D.~S.~H. {Hon}, ``{Quashing a suspected selection bias in galaxy samples having dynamically measured supermassive black holes},'' \href{http://dx.doi.org/10.1093/mnras/stac2902}{{\em \mnras} {\bfseries 518} no.~1, (Jan., 2023) 1352--1360}, \href{http://arxiv.org/abs/2210.02641}{{\ttfamily arXiv:2210.02641 [astro-ph.GA]}}.

\bibitem{MASSIVE_2014}
C.-P. {Ma}, J.~E. {Greene}, N.~{McConnell}, R.~{Janish}, J.~P. {Blakeslee}, J.~{Thomas}, and J.~D. {Murphy}, ``{The MASSIVE Survey. I. A Volume-limited Integral-field Spectroscopic Study of the Most Massive Early-type Galaxies within 108 Mpc},'' \href{http://dx.doi.org/10.1088/0004-637X/795/2/158}{{\em \apj} {\bfseries 795} no.~2, (Nov., 2014) 158}, \href{http://arxiv.org/abs/1407.1054}{{\ttfamily arXiv:1407.1054 [astro-ph.GA]}}.

\bibitem{2006AJ....131.1163S}
M.~F. {Skrutskie}, R.~M. {Cutri}, R.~{Stiening}, M.~D. {Weinberg}, S.~{Schneider}, J.~M. {Carpenter}, {\em et~al.}, ``{The Two Micron All Sky Survey (2MASS)},'' \href{http://dx.doi.org/10.1086/498708}{{\em \aj} {\bfseries 131} no.~2, (Feb., 2006) 1163--1183}.

\bibitem{6dFGS_DR1}
D.~H. {Jones}, W.~{Saunders}, M.~{Colless}, M.~A. {Read}, Q.~A. {Parker}, F.~G. {Watson}, {\em et~al.}, ``{The 6dF Galaxy Survey: samples, observational techniques and the first data release},'' \href{http://dx.doi.org/10.1111/j.1365-2966.2004.08353.x}{{\em \mnras} {\bfseries 355} no.~3, (Dec., 2004) 747--763}, \href{http://arxiv.org/abs/astro-ph/0403501}{{\ttfamily arXiv:astro-ph/0403501 [astro-ph]}}.

\bibitem{6dFGS_DR3}
D.~H. {Jones}, M.~A. {Read}, W.~{Saunders}, M.~{Colless}, T.~{Jarrett}, Q.~A. {Parker}, {\em et~al.}, ``{The 6dF Galaxy Survey: final redshift release (DR3) and southern large-scale structures},'' \href{http://dx.doi.org/10.1111/j.1365-2966.2009.15338.x}{{\em \mnras} {\bfseries 399} no.~2, (Oct., 2009) 683--698}, \href{http://arxiv.org/abs/0903.5451}{{\ttfamily arXiv:0903.5451 [astro-ph.CO]}}.

\bibitem{6dFGSv_data}
L.~A. Campbell {\em et~al.}, ``{The 6dF Galaxy Survey: Fundamental Plane Data},'' \href{http://dx.doi.org/10.1093/mnras/stu1198}{{\em Mon. Not. Roy. Astron. Soc.} {\bfseries 443} no.~2, (2014) 1231--1251}, \href{http://arxiv.org/abs/1406.4867}{{\ttfamily arXiv:1406.4867 [astro-ph.GA]}}.

\bibitem{6dFGSv_FP}
C.~{Magoulas}, C.~M. {Springob}, M.~{Colless}, D.~H. {Jones}, L.~A. {Campbell}, J.~R. {Lucey}, {\em et~al.}, ``{The 6dF Galaxy Survey: the near-infrared Fundamental Plane of early-type galaxies},'' \href{http://dx.doi.org/10.1111/j.1365-2966.2012.21421.x}{{\em \mnras} {\bfseries 427} no.~1, (Nov., 2012) 245--273}, \href{http://arxiv.org/abs/1206.0385}{{\ttfamily arXiv:1206.0385 [astro-ph.CO]}}.

\bibitem{Planck:2018vyg}
{\bfseries Planck} Collaboration, N.~Aghanim {\em et~al.}, ``{Planck 2018 results. VI. Cosmological parameters},'' \href{http://dx.doi.org/10.1051/0004-6361/201833910}{{\em Astron. Astrophys.} {\bfseries 641} (2020) A6}, \href{http://arxiv.org/abs/1807.06209}{{\ttfamily arXiv:1807.06209 [astro-ph.CO]}}. [Erratum: Astron.Astrophys. 652, C4 (2021)].

\bibitem{R_van_den_Bosch2016}
R.~C.~E. {van den Bosch}, ``{Unification of the fundamental plane and Super Massive Black Hole Masses},'' \href{http://dx.doi.org/10.3847/0004-637X/831/2/134}{{\em \apj} {\bfseries 831} no.~2, (Nov., 2016) 134}, \href{http://arxiv.org/abs/1606.01246}{{\ttfamily arXiv:1606.01246 [astro-ph.GA]}}.

\bibitem{2024MNRAS.527..249Q}
M.~E. {Quenneville}, J.~P. {Blakeslee}, C.-P. {Ma}, J.~E. {Greene}, S.~D.~J. {Gwyn}, S.~{Ciccone}, and B.~{Nyiri}, ``{The MASSIVE survey - XVIII. Deep wide-field K-band photometry and local scaling relations for massive early-type galaxies},'' \href{http://dx.doi.org/10.1093/mnras/stad3137}{{\em \mnras} {\bfseries 527} no.~1, (Jan., 2024) 249--264}, \href{http://arxiv.org/abs/2210.08043}{{\ttfamily arXiv:2210.08043 [astro-ph.GA]}}.

\bibitem{1995MNRAS.276.1341J}
I.~{Jorgensen}, M.~{Franx}, and P.~{Kjaergaard}, ``{Spectroscopy for E and S0 galaxies in nine clusters},'' \href{http://dx.doi.org/10.1093/mnras/276.4.1341}{{\em \mnras} {\bfseries 276} no.~4, (Oct., 1995) 1341--1364}.

\bibitem{kcorrection_2010}
I.~V. {Chilingarian}, A.-L. {Melchior}, and I.~Y. {Zolotukhin}, ``{Analytical approximations of K-corrections in optical and near-infrared bands},'' \href{http://dx.doi.org/10.1111/j.1365-2966.2010.16506.x}{{\em \mnras} {\bfseries 405} no.~3, (July, 2010) 1409--1420}, \href{http://arxiv.org/abs/1002.2360}{{\ttfamily arXiv:1002.2360 [astro-ph.IM]}}.

\bibitem{Schombert_Smith_2012}
J.~Schombert and A.~K. Smith, ``The Structure of Galaxies I: Surface Photometry Techniques,'' \href{http://dx.doi.org/10.1071/AS11059}{{\em Publications of the Astronomical Society of Australia} {\bfseries 29} no.~2, (2012) 174–192}.

\bibitem{2006MNRAS.366.1126C}
M.~{Cappellari}, R.~{Bacon}, M.~{Bureau}, M.~C. {Damen}, R.~L. {Davies}, P.~T. {de Zeeuw}, {\em et~al.}, ``{The SAURON project - IV. The mass-to-light ratio, the virial mass estimator and the Fundamental Plane of elliptical and lenticular galaxies},'' \href{http://dx.doi.org/10.1111/j.1365-2966.2005.09981.x}{{\em \mnras} {\bfseries 366} no.~4, (Mar., 2006) 1126--1150}, \href{http://arxiv.org/abs/astro-ph/0505042}{{\ttfamily arXiv:astro-ph/0505042 [astro-ph]}}.

\bibitem{2017A&A...597A..48F}
J.~{Falc{\'o}n-Barroso}, M.~{Lyubenova}, G.~{van de Ven}, J.~{Mendez-Abreu}, J.~A.~L. {Aguerri}, B.~{Garc{\'\i}a-Lorenzo}, {\em et~al.}, ``{Stellar kinematics across the Hubble sequence in the CALIFA survey: general properties and aperture corrections},'' \href{http://dx.doi.org/10.1051/0004-6361/201628625}{{\em \aap} {\bfseries 597} (Jan., 2017) A48}, \href{http://arxiv.org/abs/1609.06446}{{\ttfamily arXiv:1609.06446 [astro-ph.GA]}}.

\bibitem{2018MNRAS.473.5446V}
M.~{Veale}, C.-P. {Ma}, J.~E. {Greene}, J.~{Thomas}, J.~P. {Blakeslee}, J.~L. {Walsh}, and J.~{Ito}, ``{The MASSIVE survey - VIII. Stellar velocity dispersion profiles and environmental dependence of early-type galaxies},'' \href{http://dx.doi.org/10.1093/mnras/stx2717}{{\em \mnras} {\bfseries 473} no.~4, (Feb., 2018) 5446--5467}, \href{http://arxiv.org/abs/1708.00870}{{\ttfamily arXiv:1708.00870 [astro-ph.GA]}}.

\bibitem{2023RAA....23h5001Z}
K.~{Zhu}, R.~{Li}, X.~{Cao}, S.~{Lu}, M.~{Cappellari}, and S.~{Mao}, ``{Velocity Dispersion {\ensuremath{\sigma}} $_{aper}$ Aperture Corrections as a Function of Galaxy Properties from Integral-field Stellar Kinematics of 10,000 MaNGA Galaxies},'' \href{http://dx.doi.org/10.1088/1674-4527/acd58a}{{\em Research in Astronomy and Astrophysics} {\bfseries 23} no.~8, (Aug., 2023) 085001}, \href{http://arxiv.org/abs/2307.12251}{{\ttfamily arXiv:2307.12251 [astro-ph.GA]}}.

\bibitem{Mandel:2018mve}
I.~Mandel, W.~M. Farr, and J.~R. Gair, ``{Extracting distribution parameters from multiple uncertain observations with selection biases},'' \href{http://dx.doi.org/10.1093/mnras/stz896}{{\em Mon. Not. Roy. Astron. Soc.} {\bfseries 486} no.~1, (2019) 1086--1093}, \href{http://arxiv.org/abs/1809.02063}{{\ttfamily arXiv:1809.02063 [physics.data-an]}}.

\bibitem{SDSS_vdisp_DR8}
H.~{Aihara}, C.~{Allende Prieto}, D.~{An}, S.~F. {Anderson}, {\'E}.~{Aubourg}, E.~{Balbinot}, {\em et~al.}, ``{The Eighth Data Release of the Sloan Digital Sky Survey: First Data from SDSS-III},'' \href{http://dx.doi.org/10.1088/0067-0049/193/2/29}{{\em \apjs} {\bfseries 193} no.~2, (Apr., 2011) 29}, \href{http://arxiv.org/abs/1101.1559}{{\ttfamily arXiv:1101.1559 [astro-ph.IM]}}.

\bibitem{pahre_1999}
M.~A. {Pahre}, ``{Near-infrared Imaging of Early-Type Galaxies. II. Global Photometric Parameters},'' \href{http://dx.doi.org/10.1086/313249}{{\em \apjs} {\bfseries 124} no.~1, (Sept., 1999) 127--169}.

\bibitem{zahid_2016}
H.~J. {Zahid}, M.~J. {Geller}, D.~G. {Fabricant}, and H.~S. {Hwang}, ``{The Scaling of Stellar Mass and Central Stellar Velocity Dispersion for Quiescent Galaxies at z<0.7},'' \href{http://dx.doi.org/10.3847/0004-637X/832/2/203}{{\em \apj} {\bfseries 832} no.~2, (Dec., 2016) 203}, \href{http://arxiv.org/abs/1607.04275}{{\ttfamily arXiv:1607.04275 [astro-ph.GA]}}.

\bibitem{shen_2020}
X.~Shen, P.~F. Hopkins, C.-A. Faucher-Gigu\`ere, D.~M. Alexander, G.~T. Richards, N.~P. Ross, and R.~C. Hickox, ``{The bolometric quasar luminosity function at z = 0\textendash{}7},'' \href{http://dx.doi.org/10.1093/mnras/staa1381}{{\em Mon. Not. Roy. Astron. Soc.} {\bfseries 495} no.~3, (2020) 3252--3275}, \href{http://arxiv.org/abs/2001.02696}{{\ttfamily arXiv:2001.02696 [astro-ph.GA]}}.

\bibitem{2013ApJ...769L...5K}
J.~{Kormendy} and R.~{Bender}, ``{The Lvprop{\ensuremath{\sigma}}$^{8}$ Correlation for Elliptical Galaxies with Cores: Relation with Black Hole Mass},'' \href{http://dx.doi.org/10.1088/2041-8205/769/1/L5}{{\em \apjl} {\bfseries 769} no.~1, (May, 2013) L5}.

\bibitem{Thomas:2013bga}
J.~Thomas, R.~P. Saglia, R.~Bender, P.~Erwin, and M.~Fabricius, ``{The Dynamical Fingerprint of Core Scouring in Massive Elliptical Galaxies},'' \href{http://dx.doi.org/10.1088/0004-637X/782/1/39}{{\em Astrophys. J.} {\bfseries 782} no.~1, (2014) 39}, \href{http://arxiv.org/abs/1311.3783}{{\ttfamily arXiv:1311.3783 [astro-ph.GA]}}.

\bibitem{2012MNRAS.424..224H}
J.~{Hlavacek-Larrondo}, A.~C. {Fabian}, A.~C. {Edge}, and M.~T. {Hogan}, ``{On the hunt for ultramassive black holes in brightest cluster galaxies},'' \href{http://dx.doi.org/10.1111/j.1365-2966.2012.21187.x}{{\em \mnras} {\bfseries 424} no.~1, (July, 2012) 224--231}, \href{http://arxiv.org/abs/1204.5759}{{\ttfamily arXiv:1204.5759 [astro-ph.CO]}}.

\bibitem{2006MNRAS.365.1082W}
J.~S.~B. {Wyithe}, ``{A log-quadratic relation between the nuclear black hole masses and velocity dispersions of galaxies},'' \href{http://dx.doi.org/10.1111/j.1365-2966.2005.09721.x}{{\em \mnras} {\bfseries 365} no.~4, (Feb., 2006) 1082--1098}, \href{http://arxiv.org/abs/astro-ph/0503435}{{\ttfamily arXiv:astro-ph/0503435 [astro-ph]}}.

\bibitem{2009ApJ...698..198G}
K.~{G{\"u}ltekin}, D.~O. {Richstone}, K.~{Gebhardt}, T.~R. {Lauer}, S.~{Tremaine}, M.~C. {Aller}, {\em et~al.}, ``{The M-{\ensuremath{\sigma}} and M-L Relations in Galactic Bulges, and Determinations of Their Intrinsic Scatter},'' \href{http://dx.doi.org/10.1088/0004-637X/698/1/198}{{\em \apj} {\bfseries 698} no.~1, (June, 2009) 198--221}, \href{http://arxiv.org/abs/0903.4897}{{\ttfamily arXiv:0903.4897 [astro-ph.GA]}}.

\end{thebibliography}\endgroup
\bibliographystyle{utcaps}

\end{document}